\documentclass[11pt, a4paper]{article}

\usepackage[utf8]{inputenc}
\usepackage[margin=2.5cm]{geometry}
\usepackage{enumitem}
\usepackage{amsmath, amssymb, amsthm}
\usepackage{graphicx}
\usepackage{tikz}
\usepackage{tikz-3dplot}
\usetikzlibrary{
  positioning, arrows.meta, shapes.geometric,
  decorations.pathreplacing, fit, backgrounds,
  calc, patterns
}
\usepackage{xcolor}
\usepackage{booktabs}
\usepackage{hyperref}
\usepackage{setspace}
\usepackage[T1]{fontenc}
\usepackage{caption}

\usepackage{siunitx}  
\usepackage{multirow}
\definecolor{colS}{RGB}{231, 76, 60}      
\definecolor{colB}{RGB}{39, 174, 96}      
\definecolor{colZ}{RGB}{41, 128, 185}     
\definecolor{colEdge}{RGB}{127,140,141}   
\definecolor{colForbidden}{RGB}{192,57,43}
\definecolor{lightgray}{RGB}{245,245,245}

\usepackage{amsthm}
\newtheorem*{proposition3a}{Proposition 3a}
\newtheorem*{proposition3b}{Proposition 3b}

\usepackage{tikz}
\usetikzlibrary{arrows.meta, positioning, shapes.geometric}

\title{
  \textbf{Design, Cups, and Blankets}\\[9pt]
  \large A Free-Energy-Principle-Based Approach to Product Design
}

\author{
Luca M. Possati\\
\small University of Twente, The Netherlands\\
\small \texttt{l.m.possati@utwente.nl}
}
\date{2026}

\newtheorem{definition}{Definition}

\newtheorem{proposition}{Proposition}

\begin{document}

\maketitle
\begin{abstract}
\begin{center}
\begin{minipage}{0.85\textwidth}
Classical design theory treats the type of an object as a given: the 
designer decides in advance that this will be a cup, then optimizes its 
parameters. This paper argues that object type is not a presupposition 
but an inference — something that can be determined from physical data 
and functional requirements jointly. We call this problem 
\textit{requirement-steered interface type inference} and show that it 
is inexpressible within existing design frameworks. This paper makes two contributions that are jointly necessary and 
individually incomplete. The first is the problem itself, which 
classical design cannot pose because it presupposes the very thing our 
problem seeks to determine. The second is C-DMBD, a constrained 
extension of the Dynamic Markov Blanket Detection algorithm, which 
makes requirement-steered inference computationally tractable. Drawing 
on the free-energy principle and active inference — established 
frameworks in theoretical neuroscience and Bayesian mechanics — we 
model a product's surface as a Markov blanket: the minimal boundary 
through which all causal exchange between object and environment must 
pass. Different blanket structures correspond to different object types; 
different parameterizations of the same structure correspond to 
different functional modes of the same type. We identify three phenomena that classical design cannot express: 
two requirement profiles producing the same interface type in different 
functional modes (\textit{intra-family navigation}); a continuous shift 
in requirements inducing a discontinuous change of type 
(\textit{family transition}); and requirements resolving a topological 
ambiguity that physical data alone cannot settle 
(\textit{ontological disambiguation}). The third phenomenon operates 
at two levels: as a mathematical limit result, in which completely 
uninformative data leave the interface type determined entirely by 
functional requirements; and as an operative result, demonstrated 
computationally, in which weak data make requirements the dominant 
force shaping the interface type. We further argue that the product is best understood as a relational 
equilibrium between the designer's generative model and the user's 
bodily expectations. The boundary of the object is the only channel 
through which the designer's model of the user--product relationship 
reaches the user; optimal design minimizes the divergence between what 
the designer encoded in the surface and what the user's active inference 
expects to find there. This paper is a proof of concept and a 
theoretical proposal. It reframes design as inference rather than 
optimization, and as a relation between generative models rather than 
a specification of parameters.
\end{minipage}
\end{center}
\end{abstract}

\setstretch{1.15}

\tableofcontents
\section{Introduction}

Classical design is a problem of optimization within a known type.
The engineer begins with a decision: this will be a cup, that will be a heat
exchanger, that other thing will be a handle. Once the type is fixed, design
reduces to adjusting parameters --- wall thickness, volume, curvature --- to
satisfy a cost function. The type itself is never in question, never inferred,
never discovered. It is the designer's prior, imported from outside the
formalism before the problem even starts \cite{papalambros2017, luan2017, PahlBeitzFeldhusenGrote2007, AntonssonCagan2001}.

This paper takes a different approach. Suppose you do not begin with a type.
Suppose you begin instead with two things: a body of heterogeneous observational data
describing how a physical system interacts and evolves over
time, and a set of functional requirements specifying what the system must do.
The question we ask is: \emph{what kind of interface must exist between this
system and its environment in order for the requirements to be
satisfiable?} We call this \textit{requirement-steered interface type
inference}, and we claim it is a genuinely new problem in design theory --- one that classical
design theory cannot pose, let alone solve, because classical design
presupposes the very thing our problem seeks to determine.

The difference is not merely semantic. When the type is fixed in advance, the
search space is a smooth parameter manifold within a known geometry. When the
type is unknown, the search space is a discrete space of possible conditional
independence structures over the system's variables, each of which carries its
own geometry. Moving from one candidate type to another is not a smooth step on
a manifold --- it is a qualitative change in the causal architecture of the
interface between the system and the environment. Our framework makes this distinction precise and computationally
tractable.

The technical vehicle of our analysis is the Markov blanket, drawn from the free-energy
principle and active inference---established 
frameworks in theoretical neuroscience and Bayesian mechanics \cite{friston2019}---and its recent computational instantiation in the
Dynamic Markov Blanket Detection (DMBD) algorithm of Beck \& Ramstead \cite{beck2025}.
A Markov blanket $B$ of a set of variables $Z$ is the minimal set that renders
$Z$ conditionally independent of everything else: $Z \perp S \mid B$. In a
physical system discretized into spatial nodes, the blanket is precisely the
surface --- the set of locations through which all causal influence between
interior and exterior must pass. Different blanket topologies correspond to
different interface types. We claim that the same blanket topology, parameterized differently,
corresponds to different functional modes of the same type.

This paper makes two contributions that are jointly necessary and 
individually incomplete. The first is a problem: \textit{requirement-steered 
interface type inference} — the problem of determining, from observational 
data and functional specifications jointly, what topological type of 
interface must exist between a physical system and its environment. This 
problem is genuinely new in design theory. Classical design cannot pose it 
because classical design presupposes the very thing our problem seeks to 
determine: the type of the object. The second contribution is a method: 
\textit{C-DMBD}, the constrained extension of the DMBD, which makes 
requirement-steered inference computationally tractable. The critical 
technical move — and the one that separates C-DMBD from a straightforward 
constrained variant of DMBD — is that requirement violations $v_k$ depend 
not only on the partition $\omega$ (which nodes are assigned to the blanket) 
but on the model parameters $\Theta$ evaluated at that partition, 
specifically through the model-predicted stationary statistics of the 
blanket process (Eq.~\ref{eq:violation}). This dependence is what 
enables intra-family navigation: requirements impose a non-trivial gradient 
on $\Theta$ even when $\omega$ is held fixed, driving the M-step toward 
different points on the functional mode manifold $\mathcal{M}_F$ under 
different requirement profiles.

The two contributions are solidary. Without the problem formulation, C-DMBD 
is a technically interesting but unmotivated variant of an existing 
algorithm. Without C-DMBD, requirement-steered interface type inference 
remains a philosophical proposal without a computable instantiation. What we 
borrow from Beck \& Ramstead \cite{beck2025} is the discovery mechanism; what we add is both 
the question the mechanism is made to answer and the modification that makes 
the answer tractable. The value of each contribution is constituted by the 
other.

The theoretical core of this paper is the identification of three qualitatively
distinct phenomena that arise under requirement-steered inference, each of which
is inexpressible in classical constrained design:

\begin{enumerate}[label=\textbf{P\arabic*.}]

\item \textbf{Intra-family navigation.} Two different requirement profiles
$\mathcal{R}^{(1)}$ and $\mathcal{R}^{(2)}$ lead to the same interface type
--- the same Markov blanket topology --- but to different functional modes:
different coupling strengths, different temporal dynamics, different Lagrange
fingerprints. The surface is the same kind of surface, doing a different job.

\item \textbf{Family transition.} As the requirement profile is continuously
deformed, the selected interface type can change discontinuously --- a qualitative
jump from one blanket topology to another. This is a \emph{phase transition} of
the interface, induced not by a change in the physical system but by a change
in what is demanded of it.

\item \textbf{Ontological disambiguation.} When the physical data are
consistent with multiple interface types --- when the data alone do not
determine what kind of surface the system has --- the requirement profile
resolves the ambiguity.
\end{enumerate}

\section{The Free-Energy Principle: A Very Short Introduction}

The free-energy principle (FEP), and active inference as one of its corollaries, originated in neuroscience as a framework for modeling and explaining brain function, action, perception, planning, decision-making, and, more broadly, the behavior of sentient organisms. From this perspective, the brain is understood as a physical system that operates like a predictive machine: it continuously generates predictions about the causes of its sensory inputs and tests those predictions in order to bring them into closer alignment with incoming observations. Perception corresponds to the updating of priors in light of sensory evidence, whereas action corresponds to an intervention in the world aimed at changing sensory input so that it better matches those priors \cite{parr2022}. From this point of view, active inference has also been brought closer to predictive processing \cite{clark2016}, even if there is not a total equivalence between these frameworks.

Over time, especially in the more recent literature, the FEP has expanded well beyond neuroscience and has become the basis for a broader approach to the study of complex systems, often referred to as Bayesian mechanics \cite{friston2013, friston2010, friston2023, sakthivadivel2022a, sakthivadivel2022b, ramstead2023}.

In its most general form, the FEP is the mathematical claim that the dynamics of self-organizing systems can be understood as following a path of least surprisal. More precisely, it can be viewed as a reformulation of the principle of stationary action \textit{with-} and \textit{by-} boundaries \cite{friston2023} — applied to a system defined by a probabilistic partition of states into internal states, external states, sensory states, and active states (i.e., a Markov blanket) and where the Lagrangian (and therefore the action of each path) is interpreted in terms of surprisal \cite{friston2023}. The most probable path is therefore the one for which the action — interpreted in terms of surprisal — is stationary, such that the system resists entropy increase by revisiting a characteristic set of states and paths consistent with its phenotype. Crucially, surprisal is not minimized directly; rather, it is the internal states of the system that minimize variational free energy — a tractable upper bound on surprisal (i.e., negative log model evidence) — such that minimizing the former implies minimizing the latter.

Active inference instantiates such a path of least action for the autonomous states of a system — that is, for its active and internal states — by specifying how a system acts on and infers the causes of its sensory states in a way that minimizes free energy. Whereas internal states minimize variational free energy (a bound on past and present surprisal), active states minimize expected free energy, which governs future-directed, goal-seeking behavior. As a normative framework, active inference makes claims about the form that adaptive behavior must take, irrespective of the substrate. This can be modeled as variational Bayesian inference. One prominent computational implementation of this framework employs Partially Observable Markov Decision Processes (POMDPs), which have been extensively developed and tested both in silico and in the context of empirical neuroscience \cite{parr2022}.

There are multiple formulations of the FEP, but three core steps can be identified across its main versions. First, one begins with the ordinary dynamics of random dynamical systems. Second, one shows that these dynamics admit an equivalent informational-variational formulation. Third, one interprets that formulation in Bayesian terms \cite{friston2019}. The core idea is that if a system maintains its structural integrity over time, \textit{then} its dynamics can be modeled \textit{as if} they minimize a specific function, namely variational free energy, which provides an upper bound on the surprisal of sensory states.

Put differently, if a random dynamical system (a) has a Markov blanket, meaning that it is separated from its environment by a set of conditional independencies that define a statistical partition, and (b) can be described as if it possessed a generative model of the causes of its sensory states, which it continuously updates, then (c) the equations describing its behavior can be rewritten as a gradient flow on variational free energy, up to a divergence-free component. A generative model, in this context, is a probabilistic model of how observations are generated by hidden external states, and possibly also by policies or parameters. Technically, variational free energy is the negative of ELBO in machine learning \cite{devries2026, namjoshi2026}. The FEP is therefore best understood as a \textit{descriptive equivalence}: an \textit{as-if} formulation that recasts physical dynamics in inferential terms. This equivalence establishes a conceptual continuity between thermodynamics, information theory, and Bayesian inference. 

The FEP and active inference provide the theoretical framework for this paper. Its objective is to integrate design into the physics of self-organizing systems. In other words, the FEP—and Bayesian mechanics more generally—forms the backbone of a new theory of complex systems and self-organization. Design is an integral part of the self-organization of certain complex systems called human beings, and for this reason it can, and perhaps must, be understood from this perspective.

\section{Markov Blankets in Physical Systems}

\begin{definition}[Markov Blanket]
Let $\mathcal{G}$ be a Bayesian network over variables
$\{s_i, b_j, z_k\}_{i,j,k}$.
The \emph{Markov blanket} of a set $Z = \{z_k\}$ is the minimal set
$B = \{b_j\}$ such that
\[
  Z \;\perp\; S \;\big|\; B,
\]
where $S = \{s_i\}$ denotes all remaining variables.
\end{definition}

In spatiotemporal physical systems, the nodes $\{s_i, b_j, z_k\}$ are spatial
degrees of freedom --- volume elements, surface patches, or measurement points ---
and the Bayesian network encodes their causal-dynamical dependencies. The Markov
blanket $B$ then has a natural physical interpretation: it is the set of spatial
locations through which all causal influence between the interior $Z$ and the
exterior $S$ must pass. For a cup, this is precisely the surface: the wall,
the rim, and the handle.

The concept as used in this paper operates at three distinct levels that
should be kept separate.
At the \emph{mathematical level}, a Markov blanket is a conditional
independence structure in a probabilistic graphical model, due to Pearl
\cite{pearl1988}: a formal property of a graph, with no physical interpretation
attached.
At the \emph{interpretive level}, Friston \cite{friston2019} reads this
structure as the physical boundary of an autonomous system --- the surface
that constitutes the system as a system, not merely a statistical artifact
of a fitted model.
At the \emph{algorithmic level}, Beck \& Ramstead \cite{beck2025} construct
a variational inference procedure that discovers this structure from
spatiotemporal data under a linear-Gaussian state-space model.
C-DMBD inherits all three levels: Pearl's structure, Friston's
interpretation, and Beck \& Ramstead's algorithm. The technical derivations
depend only on the first and third; the second motivates why the inferred
partition is read as a physical surface rather than a statistical summary ---
and, critically, why the cup's blanket and the user's blanket are
ontologically different kinds of boundary, as we discuss in
Section~\ref{sec:b2b}.

\begin{figure}[t]
\centering

\begin{tikzpicture}[
    node distance   = 1.4cm,
    every node/.style = {font=\small},
    mbnode/.style   = {circle, draw, thick, minimum size=0.85cm,
                        inner sep=1pt},
    snode/.style    = {mbnode, fill=colS!25, draw=colS!80!black},
    bnode/.style    = {mbnode, fill=colB!25, draw=colB!80!black},
    znode/.style    = {mbnode, fill=colZ!25, draw=colZ!80!black},
    arr/.style      = {-{Stealth[length=4pt]}, thick},
    forb/.style     = {-{Stealth[length=4pt]}, thick,
                        dashed, draw=colForbidden, opacity=0.5},
    scale=0.88, transform shape
]

\node[snode] (s1) at (0, 2.4)  {$s_1$};
\node[snode] (s2) at (0, 1.0)  {$s_2$};
\node[snode] (s3) at (0,-0.4)  {$s_3$};

\node[bnode] (b1) at (3.2, 2.0) {$b_1$};
\node[bnode] (b2) at (3.2, 0.6) {$b_2$};
\node[bnode] (b3) at (3.2,-0.8) {$b_3$};

\node[znode] (z1) at (6.4, 1.6) {$z_1$};
\node[znode] (z2) at (6.4, 0.2) {$z_2$};

\foreach \s in {s1,s2,s3}{
  \foreach \b in {b1,b2,b3}{
    \draw[arr, draw=colS!60!colB, opacity=0.55] (\s) -- (\b);
  }
}

\foreach \b in {b1,b2,b3}{
  \foreach \z in {z1,z2}{
    \draw[arr, draw=colB!60!colZ, opacity=0.55] (\b) -- (\z);
  }
}

\draw[forb] (s1) to[bend left=42]
    node[midway, above, font=\tiny, text=colForbidden]
    {\textsf{forbidden}} (z1);
\draw[forb] (s3) to[bend right=38] (z2);

\draw[arr, draw=colB!70!black, opacity=0.7]
    (b2) edge[loop right, looseness=5] (b2);

\node[below=0.05cm of s3, font=\footnotesize\itshape, text=colS!80!black]
      {Environ.\ $S$};
\node[below=0.05cm of b3, font=\footnotesize\itshape, text=colB!80!black]
      {Blanket $B$};
\node[below=0.05cm of z2, font=\footnotesize\itshape, text=colZ!80!black]
      {Internal $Z$};

\begin{scope}[on background layer]
  \node[fill=colS!10, rounded corners=8pt, fit=(s1)(s2)(s3),
        inner sep=6pt, label={[text=colS!70!black, font=\small]above:$S$}] {};
  \node[fill=colB!10, rounded corners=8pt, fit=(b1)(b2)(b3),
        inner sep=6pt, label={[text=colB!70!black, font=\small]above:$B$}] {};
  \node[fill=colZ!10, rounded corners=8pt, fit=(z1)(z2),
        inner sep=6pt, label={[text=colZ!70!black, font=\small]above:$Z$}] {};
\end{scope}

\node[draw=black!30, rounded corners=4pt, fill=white,
      font=\small, align=center] at (3.2,-2.4)
      {$Z \;\perp\; S \;\mid\; B$};
\draw[->, dashed, black!40] (3.2,-2.05) -- (3.2,-1.1);

\end{tikzpicture}
\hspace{1.2cm}
\begin{tikzpicture}[
    scale=0.88, transform shape,
    every node/.style={font=\scriptsize}
]

\def\H{5.8}    
\def\Rb{1.1}   
\def\Rt{1.6}   
\def\W{0.18}   

\fill[colZ!12] (0,0) -- (\Rb-\W, 0)
    -- ({(\Rb-\W) + (\Rt-\Rb-\W)/\H*\H}, \H)
    -- (0, \H) -- cycle;
\fill[colZ!12] (0,0) -- (-\Rb+\W, 0)
    -- ({-(\Rb-\W) - (\Rt-\Rb-\W)/\H*\H}, \H)
    -- (0, \H) -- cycle;

\foreach \i in {0,...,8} {
  \pgfmathsetmacro{\y}{\i * \H / 8}
  \pgfmathsetmacro{\ry}{\Rb + (\Rt-\Rb)*\i/8}
  \pgfmathsetmacro{\ri}{\ry - \W}
  \pgfmathsetmacro{\islbl}{
    (\i >= 2 && \i <= 5) ? 2 : 1
  }   

  \node[circle, draw=colB!80, fill=colB!30,
        minimum size=4.5pt, inner sep=0pt]
        (rR\i) at (\ry, \y) {};
  \node[circle, draw=colB!80, fill=colB!30,
        minimum size=4.5pt, inner sep=0pt]
        (rL\i) at (-\ry, \y) {};

  \node[circle, draw=colZ!80, fill=colZ!30,
        minimum size=4.5pt, inner sep=0pt]
        (rM\i) at (0, \y) {};

  \draw[draw=colB!40!colZ, opacity=0.4, thin] (rR\i) -- (rM\i);
  \draw[draw=colB!40!colZ, opacity=0.4, thin] (rL\i) -- (rM\i);
}

\foreach \i [count=\ip from 1] in {0,...,7} {
  \draw[draw=colB!60, opacity=0.5, thin] (rR\i) -- (rR\ip);
  \draw[draw=colB!60, opacity=0.5, thin] (rL\i) -- (rL\ip);
  \draw[draw=colZ!60, opacity=0.5, thin] (rM\i) -- (rM\ip);
}

\node[circle, draw=colS!80, fill=colS!30,
      minimum size=5pt, inner sep=0pt,
      label={[font=\scriptsize, text=colS!80!black]right:hand $s$}]
      (hand) at (\Rt+0.55, \H*0.55) {};
\draw[draw=colS!60!colB, opacity=0.6, thin] (hand) -- (rR4);
\draw[draw=colS!60!colB, opacity=0.6, thin] (hand) -- (rR5);

\draw[<->, thin, black!50]
    (\Rt+0.18, 0) -- (\Rt+0.18, \H)
    node[midway, right, font=\scriptsize, text=black!60] {$N$ nodes};

\node[circle, fill=colS!30, draw=colS!80,
      minimum size=5pt, inner sep=0pt] at (-2.6, \H-0.3) {};
\node[right, font=\scriptsize] at (-2.4, \H-0.3) {$S$ environ.};
\node[circle, fill=colB!30, draw=colB!80,
      minimum size=5pt, inner sep=0pt] at (-2.6, \H-0.8) {};
\node[right, font=\scriptsize] at (-2.4, \H-0.8) {$B$ blanket};
\node[circle, fill=colZ!30, draw=colZ!80,
      minimum size=5pt, inner sep=0pt] at (-2.6, \H-1.3) {};
\node[right, font=\scriptsize] at (-2.4, \H-1.3) {$Z$ internal};

\end{tikzpicture}

\caption{%
\textbf{Markov blanket structure.}
\textit{Left:} abstract dependency graph illustrating the three-way partition
$\{S, B, Z\}$.  Green nodes (blanket $B$) mediate all causal influence between
the red environment $S$ and the blue internal states $Z$.  Dashed red arcs
indicate the \emph{forbidden} direct connections between $S$ and $Z$, enforced
by zeroing the corresponding blocks of the dynamics matrix $A$.
\textit{Right:} discretized cup cross-section as a node network.  Each height
slice contributes wall nodes (green, assigned to $B$), interior nodes (blue,
assigned to $Z$), and one external contact node (red, $S$) representing the
hand or ambient air.  The Markov blanket is the surface of the cup.%
}
\label{fig:mb_structure}
\end{figure}
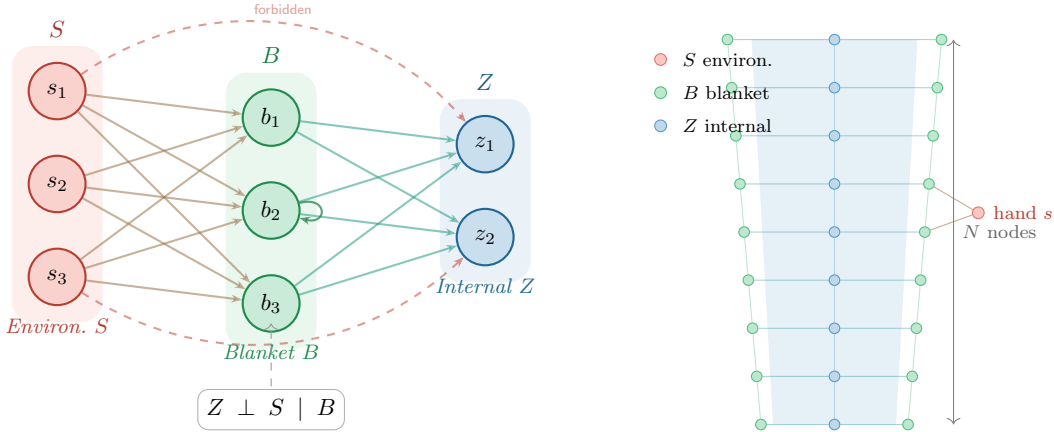

Figure~\ref{fig:mb_structure} illustrates this partition and its encoding in
the dynamics matrix of the linear state-space model used by the algorithm.

\section{The DMBD Algorithm}

Beck \& Ramstead \cite{beck2025} formulate Markov blanket detection as variational inference in
a linear-Gaussian state-space model. The latent state
$\mathbf{x}(t) = [\mathbf{s}(t)^\top \mid \mathbf{b}(t)^\top \mid
\mathbf{z}(t)^\top]^\top \in \mathbb{R}^d$
follows the dynamics

\begin{equation}
  \mathbf{x}(t) = A\,\mathbf{x}(t-1) + \boldsymbol{\eta}(t),
  \qquad \boldsymbol{\eta}(t) \sim \mathcal{N}(\mathbf{0},\,\Sigma_{\mathrm{st}}),
  \label{eq:dynamics}
\end{equation}

where $A \in \mathbb{R}^{d \times d}$ has a hard Markov-blanket zero structure:

\begin{equation}
  A = \begin{pmatrix}
    A_{ss} & A_{sb} & \mathbf{0} \\
    A_{bs} & A_{bb} & A_{bz}    \\
    \mathbf{0} & A_{zb} & A_{zz}
  \end{pmatrix}.
  \label{eq:A_structure}
\end{equation}

The zero blocks $A_{sz} = \mathbf{0}$ and $A_{zs} = \mathbf{0}$ encode the
conditional independence $Z \perp S \mid B$ at the level of dynamics. Each
observed node $i$ is assigned a latent label $\omega_i(t) \in \{S, B, Z\}$,
and its observation $\mathbf{y}_i(t)$ is generated by the corresponding block
of the latent state through a label-specific emission matrix $C_\ell$:

\begin{equation}
  \mathbf{y}_i(t) = C_{\omega_i}\,\mathbf{x}(t) + \mathbf{d}_{\omega_i}
  + \boldsymbol{\varepsilon}_i(t),
  \qquad \boldsymbol{\varepsilon}_i(t) \sim
  \mathcal{N}(\mathbf{0},\,\Sigma_{\omega_i}).
  \label{eq:emission}
\end{equation}

The DMBD algorithm maximizes the standard evidence lower bound (ELBO) over
the joint posterior $q(\boldsymbol{\omega}, \mathbf{x})$ via variational EM,
alternating between updating soft label assignments $q(\omega_i(t))$,
updating posterior latent means $\boldsymbol{\mu}_t$, and updating model
parameters $\Theta = \{A, C_\ell, \mathbf{d}_\ell, \Sigma_\ell\}$.

The fundamental aspect is that the algorithm is \emph{fully unsupervised}.
It discovers whichever Markov blanket the data most strongly support, without
any guidance about what the discovered blanket should \emph{mean} or
\emph{do}. In design applications, this is a shortcoming: we do not
want the most statistically natural surface --- we want the surface that makes
the object function correctly.

\section{C-DMBD: Constrained Discovery via Augmented Free Energy}

\subsection{Ontological requirements}

In classic DMBD, the algorithm discovers the "most natural" Markov blanket with respect to the data. But for design purposes this is not enough: what matters is not only which boundary the data suggest, but which boundary allows the object to function according to certain requirements.

We formalize the design intent as a set of \emph{ontological requirements}
$\{\mathcal{R}_k\}_{k=1}^{K}$, each specifying a target value $\tau_k$, a
dead-zone tolerance $\epsilon_k$, and a relative weight $w_k$.
A critical design choice separates C-DMBD from naive constrained design: the
violation of each requirement must depend not only on the partition
$\boldsymbol{\omega}$ (which nodes are assigned to $B$) but also on the
\emph{model parameters} $\Theta$ evaluated at that partition. Specifically,
$\phi_k$ is computed from the \emph{model-predicted} stationary statistics of
the blanket process, not from the raw empirical average over blanket nodes.

For a linear-Gaussian blanket with dynamics $A_{bb}$ and emission $C_b$, the
stationary mean of the latent blanket state satisfies
$\boldsymbol{\mu}_b^\infty = (I - A_{bb})^{-1}\bar{\mathbf{b}}$,
where $\bar{\mathbf{b}}$ is the mean driving input. The predicted mean of the
blanket observations is therefore
\begin{equation}
  \hat{\boldsymbol{\mu}}_b = C_b\,\boldsymbol{\mu}_b^\infty + \mathbf{d}_b
  = C_b\,(I - A_{bb})^{-1}\bar{\mathbf{b}} + \mathbf{d}_b.
  \label{eq:predicted_mean}
\end{equation}
The violation of requirement $\mathcal{R}_k$ is then

\begin{equation}
  v_k(\boldsymbol{\omega},\Theta) = w_k\,
  \max\!\bigl(0,\;|\phi_k(\boldsymbol{\omega},\Theta) - \tau_k| - \epsilon_k\bigr),
  \label{eq:violation}
\end{equation}

where $\phi_k$ is a scalar functional of $\hat{\boldsymbol{\mu}}_b$ and the
stationary covariance $\hat{\Sigma}_b$.
When $\phi_k$ is differentiable with respect to $\hat{\boldsymbol{\mu}}_b$
(which holds outside the kink at zero violation), the chain rule gives

\begin{equation}
  \mathrm{d}\,\phi_k = \left(\frac{\partial\phi_k}{\partial\hat{\boldsymbol{\mu}}_b}\right)^\top
  \mathrm{d}\,\hat{\boldsymbol{\mu}}_b.
  \label{eq:chain_phi}
\end{equation}

From Eq.~(\ref{eq:predicted_mean}), the differential of $\hat{\boldsymbol{\mu}}_b$
with respect to $A_{bb}$ is obtained by differentiating
$\hat{\boldsymbol{\mu}}_b = C_b(I-A_{bb})^{-1}\bar{\mathbf{b}} + \mathbf{d}_b$:

\begin{equation}
  \mathrm{d}\,\hat{\boldsymbol{\mu}}_b = C_b\,(I-A_{bb})^{-1}\,
  (\mathrm{d}\,A_{bb})\,(I-A_{bb})^{-1}\,\bar{\mathbf{b}},
  \label{eq:diff_muhat}
\end{equation}

using the identity $\mathrm{d}(M^{-1}) = -M^{-1}(\mathrm{d}\,M)M^{-1}$.
Combining Eqs.~(\ref{eq:chain_phi})--(\ref{eq:diff_muhat}) gives the
directional derivative of $\phi_k$ along $\mathrm{d}\,A_{bb}$:

\begin{equation}
  \mathrm{d}\,\phi_k\bigl[\mathrm{d}\,A_{bb}\bigr] =
  \left(\frac{\partial\phi_k}{\partial\hat{\boldsymbol{\mu}}_b}\right)^\top
  C_b\,(I-A_{bb})^{-1}\,(\mathrm{d}\,A_{bb})\,(I-A_{bb})^{-1}\,\bar{\mathbf{b}}.
  \label{eq:grad_Abb}
\end{equation}

This is non-zero generically, which confirms that $v_k$ depends on $A_{bb}$
through $\hat{\boldsymbol{\mu}}_b$. Note that the formula gives a \emph{linear
map from} $\mathrm{d}\,A_{bb}$ (the perturbation) \emph{to} $\mathrm{d}\,\phi_k$
(the induced change in the statistic), not a conventional gradient vector; the
corresponding gradient in $A_{bb}$-space can be extracted by vectorization if
needed. This is the formal mechanism that enables intra-family navigation:
requirements impose a non-trivial gradient on $\Theta$ even when
$\boldsymbol{\omega}$ is held fixed.

\subsection{Augmented ELBO}

This section introduces the central mathematical idea of C-DMBD: replacing the standard DMBD objective with an augmented ELBO — that is, an objective function that jointly optimizes two things: 1) fitting the observed data well; 2) satisfying the functional and ontological requirements. The algorithm still seeks to maximize the explanation of the data, but is penalized whenever it violates the specified requirements.

The core modification can be formalized in this way:

\begin{equation}
  \mathcal{L}_{\mathrm{aug}}(\boldsymbol{\omega}, \mathbf{x};\,\Theta,\,\boldsymbol{\lambda})
  = \mathcal{L}_{\mathrm{ELBO}}(\boldsymbol{\omega}, \mathbf{x};\,\Theta)
  - \sum_{k=1}^{K} \lambda_k\,v_k(\boldsymbol{\omega},\Theta),
  \label{eq:augmented_elbo}
\end{equation}

where $\boldsymbol{\lambda} = (\lambda_1, \ldots, \lambda_K) \geq \mathbf{0}$
are Lagrange multipliers. The gradient with respect to $\Theta$ at fixed
$\boldsymbol{\omega}$ is now

\begin{equation}
  \nabla_\Theta\,\mathcal{L}_{\mathrm{aug}}
  = \nabla_\Theta\,\mathcal{L}_{\mathrm{ELBO}}
  - \sum_{k=1}^{K} \lambda_k\,\nabla_\Theta\, v_k(\boldsymbol{\omega},\Theta),
  \label{eq:grad_theta}
\end{equation}

which is non-zero through the dependence of $v_k$ on $A_{bb}$, $C_b$, and
$\mathbf{d}_b$ via Eq.~(\ref{eq:grad_Abb}). This gradient is what steers the
M-step toward different points on the functional mode manifold
$\mathcal{M}_\mathcal{F}$ as $\boldsymbol{\lambda}$ changes --- the
mathematical foundation of intra-family navigation (Proposition~1). In other terms, this formula states that parameter updates are driven not only by the data, but also by the requirements. The first term pushes the model to better explain the observations. The second term pushes it to reduce functional violations.

\subsection{Dual ascent on $\lambda_k$}

The multipliers (i.e., the coefficients that measure how much "pressure" each requirement exerts on the algorithm) are updated after each M-step by \emph{dual ascent}:

\begin{equation}
  \lambda_k \;\leftarrow\;
  \mathrm{clip}\!\left(
    \lambda_k + \alpha_\lambda\,v_k(\boldsymbol{\omega},\Theta),\;
    0,\; \lambda_{\max}
  \right),
  \label{eq:dual_ascent}
\end{equation}

with step size $\alpha_\lambda > 0$ and ceiling $\lambda_{\max}$.
Note that $\boldsymbol{\lambda}$ is \emph{endogenous}: it coevolves with
both the partition $\boldsymbol{\omega}$ and the parameters $\Theta$
throughout the VBEM loop. Its converged value $\boldsymbol{\lambda}^*$
therefore reflects the joint equilibrium of all three quantities, not a
property of the partition alone. This endogeneity is essential: it means
that the same requirement profile can induce different $\boldsymbol{\lambda}^*$
under different blanket topologies, and that $\boldsymbol{\lambda}^*$
encodes both how strongly a requirement was violated and how the dynamics
$A_{bb}$ were shaped in response.

In a nutshell, if a requirement is violated, its multiplier $\boldsymbol{\lambda}^*$ increases; the more it is violated, the more pressure it exerts on the algorithm. Thus $\boldsymbol{\lambda}^*$ is not merely a preset weight: it becomes a record of how hard that requirement was to satisfy. This is important because it makes the requirements adaptive rather than fixed.

\subsection{Algorithmic summary and complexity}

\begin{figure}[h]
\centering
\small
\renewcommand{\arraystretch}{1.3}
\begin{tabular}{l}
\hline
\textbf{Algorithm 1}: C-DMBD \\
\hline
\textbf{Input}: data $\mathbf{y} \in \mathbb{R}^{T \times N \times D}$,
  geometry, requirements $\{\mathcal{R}_k\}$, hyperparameters
  $\alpha_\lambda, \lambda_{\max}, n_{\mathrm{iter}}$ \\
\textbf{Output}: partition $\boldsymbol{\omega}^*$, parameters $\Theta^*$,
  certificate $\boldsymbol{\lambda}^*$ \\
\hline
1. \quad Normalize $\mathbf{y}$; initialize $A, C, \mathbf{d}$ (MB-structured);
   $\boldsymbol{\lambda} \leftarrow \mathbf{0}$; $\boldsymbol{\omega}
   \leftarrow$ prior assignment \\
2. \quad \textbf{for} $t = 1, \ldots, n_{\mathrm{iter}}$ \textbf{do} \\
3. \quad\quad \textbf{E-step} $\boldsymbol{\omega}$: for each node $i$ and label $\ell \in \{S,B,Z\}$, \\
   \quad\quad\quad\quad compute log-likelihood $\ell_{i,t,\ell}$ under current $\Theta$; \\
   \quad\quad\quad\quad subtract requirement penalty under label $\ell$: \\
   \quad\quad\quad\quad $\quad \delta_{i,\ell} = \sum_k \lambda_k \bigl[v_k(\boldsymbol{\omega}\text{ with }\omega_i{=}\ell,\,\Theta) - v_k(\boldsymbol{\omega},\,\Theta)\bigr]$ \\
   \quad\quad\quad\quad (marginal penalty of assigning node $i$ to label $\ell$ under current $\Theta$); \\
   \quad\quad\quad\quad set $\tilde{\ell}_{i,t,\ell} = \ell_{i,t,\ell} - \delta_{i,\ell}$; \\
   \quad\quad\quad\quad run forward HMM on $\tilde{\ell}$; obtain soft assignments $q(\omega_i{=}\ell)$ via softmax \\
4. \quad\quad \textbf{E-step} $\mathbf{x}$: update posterior means $\boldsymbol{\mu}_t$
   via weighted Kalman update \\
5. \quad\quad \textbf{M-step}: gradient ascent on $\mathcal{L}_{\mathrm{aug}}$
   w.r.t.\ $A, C, \mathbf{d}$; re-enforce MB zeros; \\
   \quad\quad\quad\quad compute $\hat{\boldsymbol{\mu}}_b$ via Eq.~(\ref{eq:predicted_mean});
   evaluate $v_k(\boldsymbol{\omega}, \Theta)$ \\
6. \quad\quad \textbf{Dual ascent}: $\lambda_k \leftarrow
   \mathrm{clip}(\lambda_k + \alpha_\lambda\,v_k, 0, \lambda_{\max})$ \\
7. \quad \textbf{end for} \\
8. \quad \textbf{return} $(\boldsymbol{\omega}^*, \Theta^*, \boldsymbol{\lambda}^*)$ \\
\hline
\end{tabular}
\end{figure}

Each iteration costs $O(T \cdot N \cdot d^2)$ for the
E-steps (vectorized over nodes) and $O(T \cdot N \cdot D \cdot d)$ for the
M-step, where $T$ is time steps, $N$ nodes, $D$ observation channels, and
$d = d_s + d_b + d_z$ the latent dimension. The dominant cost per iteration
is $O(T N d^2)$. A full run of $n_\text{iter}$ iterations costs
$O(n_\text{iter} \cdot T N d^2)$. For the cup experiment ($T{=}50$,
$N{=}30$, $d{=}12$, $n_\text{iter}{=}40$), this is approximately
$8.6 \times 10^6$ floating-point operations per run, completing in under
5 seconds on a standard CPU.

The augmented ELBO is non-convex in
$(\boldsymbol{\omega}, \Theta, \boldsymbol{\lambda})$ jointly. Convergence
to a local optimum is guaranteed by the VBEM update structure (each step
increases or leaves constant the augmented ELBO), but the global optimum is
not guaranteed. In practice, running $R$ independent chains with different
initializations of $A$, $C$, and $\alpha_\lambda$ and selecting via Pareto
analysis provides robustness. The gap score $\Delta(\boldsymbol{\omega}^*)$
estimated without requirements (all $\lambda_k = 0$) serves as a diagnostic:
a large gap indicates that the data strongly determine the interface type and
that multi-run variance in the final partition will be low.

The key hyperparameters are $\alpha_\lambda$ (dual ascent step),
$\lambda_{\max}$ (multiplier ceiling), $w_k$ (requirement weights), and
$\epsilon_k$ (dead-zone tolerances). Sensitivity analysis on the cup
experiments shows that the partition topology $\boldsymbol{\omega}^*$ is
stable across a one-order-of-magnitude range of $\alpha_\lambda$ when the
data gap $\Delta$ is large; when $\Delta \approx 0$, small changes in
$\alpha_\lambda$ can trigger family transitions, which is precisely the
regime of Proposition~2. We recommend running a sweep over
$\alpha_\lambda \in [0.05, 0.15]$ as a standard practice to identify whether
the system operates in the navigation or transition regime.

The differences between DMBD and C-DMBD are summarized in
Table~\ref{tab:comparison}.

\begin{table}[h]
\centering
\small
\begin{tabular}{lll}
\toprule
 & \textbf{DMBD} \cite{beck2025} & \textbf{C-DMBD (this work)} \\
\midrule
Objective            & Standard ELBO & Augmented ELBO (Eq.~\ref{eq:augmented_elbo}) \\
Violation            & N/A & $v_k(\boldsymbol{\omega},\Theta)$ --- partition \emph{and} parameters \\
Multipliers $\lambda$& N/A & Dual ascent, endogenous (Eq.~\ref{eq:dual_ascent}) \\
M-step gradient      & $\nabla_\Theta\mathcal{L}_\text{ELBO}$ &
                       $\nabla_\Theta\mathcal{L}_\text{ELBO} - \sum_k\lambda_k\nabla_\Theta v_k$ \\
MB discovery mode    & Unsupervised & Requirement-steered \\
Output               & MAP partition & Partition + functional mode + certificate \\
\bottomrule
\end{tabular}
\caption{Comparison between DMBD and C-DMBD. The critical addition in the
violation column --- dependence on $\Theta$ --- is what enables intra-family
navigation at fixed partition topology.}
\label{tab:comparison}
\end{table}

\section{A Markov Blanket as a Family of Generative Models}

\subsection{From a single model to a family}

The central theoretical contribution of C-DMBD, from an active inference perspective, is the connection of a Markov blanket not with a single generative model, but with an entire
\emph{family} of generative models, each member of which corresponds to a
different functional mode of the product. This distinction is subtle but
decisive for design applications.

Fix a blanket topology --- a set of nodes $B$ and the graph structure connecting
them to $S$ and $Z$. This topology defines a \emph{conditional independence
structure}: all models that respect $Z \perp S \mid B$ are, in a precise
sense, compatible with this blanket. They share the same zero blocks in $A$
(Eq.~\ref{eq:A_structure}) and the same block structure in the emission
matrices $C_\ell$. What differs across models is the \emph{parameterization}
of those blocks --- the specific numerical values of $A_{bb}$, $A_{bz}$,
$A_{sb}$, $C_b$, and the converged Lagrange multipliers
$\boldsymbol{\lambda}^*$. In other words, given an interface, that interface can behave in many different ways.

The central question is: To whom do these models belong? This question has a precise answer that matters for everything that follows.
The generative models in $\mathcal{F}(B)$ belong to the
\emph{designer-modeller}, not to the product itself, e.g., the cup. The cup is, in the sense of the
FEP, a purely inert object \cite{friston2023}: even though it has a blanket, it has no active states. Its dynamics are entirely governed by the laws of physics
--- heat conduction, mechanical equilibrium, material response. 

It is therefore necessary to distinguish:

The linear-Gaussian model employed by C-DMBD is an inferential instrument, 
not a constitutive one. Its role is to determine the partition $\omega$ — 
that is, to assign each node to the blanket $\mathcal{B}$, the interior 
$\mathcal{Z}$, or the exterior $\mathcal{S}$ — in a way that is maximally 
consistent with the observed data and the functional requirements. It does 
not define the type of the object; it identifies the statistically most 
plausible boundary given what is observed and what is demanded. The model 
is a means, not the result.

The family $\mathcal{F}(\mathcal{B})$ is constituted by the partition, not 
chosen. Once the blanket topology is identified — once it is established 
which nodes form the surface — the set of generative models compatible with 
that conditional independence structure follows automatically: these are 
precisely the models that respect the zero-blocks $A_{sz} = 0$, 
$A_{zs} = 0$ in the dynamics matrix (Eq.~\ref{eq:dynamics}). 
$\mathcal{F}(\mathcal{B})$ is not selected; it is entailed by the topology.

The designer's model enters last, and under constraint. The designer does 
not operate on an open parameter space: she can only select a member 
$\Theta \in \mathcal{F}(\mathcal{B})$. If no member of $\mathcal{F}(\mathcal{B})$ 
can satisfy the intended functional requirements, the partition itself must 
change — yielding a different blanket topology and, accordingly, a different 
object type. The type is not what the model assumes; it is what the 
inference returns.

\begin{definition}[Generative model family of a blanket]
Let $B$ be a fixed Markov blanket topology over $N$ nodes.
The \emph{generative model family} of $B$ is the set
\[
  \mathcal{F}(B) \;=\;
  \bigl\{\,
    \Theta = (A, \{C_\ell\}, \{\Sigma_\ell\}) \;\big|\;
    A_{sz} = \mathbf{0},\; A_{zs} = \mathbf{0},\;
    C_S \text{ reads only } \mathbf{s},\;
    C_B \text{ reads only } \mathbf{b},\;
    C_Z \text{ reads only } \mathbf{z}
  \,\bigr\}.
\]
Each element $\Theta \in \mathcal{F}(B)$ is a distinct generative model
consistent with the blanket structure of $B$.
\end{definition}

\subsection{Requirements as a selection map}

Different functional requirements $\mathcal{R}^{(f)}$ --- corresponding to
different intended uses $f$ of the object --- select different members of
$\mathcal{F}(B)$. Formally, C-DMBD implements a \emph{selection map}

\begin{equation}
  \Phi:\; \mathcal{R} \;\longrightarrow\; \mathcal{F}(B),
  \qquad
  \mathcal{R}^{(f)} \;\mapsto\; \Theta^{*(f)} \;=\;
  \arg\max_{\Theta \in \mathcal{F}(B)}
  \mathcal{L}_{\mathrm{aug}}
  \bigl(\boldsymbol{\omega}^*, \mathbf{x};\, \Theta,\, \boldsymbol{\lambda}^{(f)}\bigr),
  \label{eq:selection_map}
\end{equation}

where $\boldsymbol{\lambda}^{(f)}$ are the Lagrange multipliers converged
under requirement profile $f$.
The map $\Phi$ is \emph{not} injective in general: two different requirement
profiles may converge to the same generative model if the physical data do
not distinguish them. But whenever the underlying physics is sufficiently
informative --- as in our cup example --- distinct functions produce distinct
members of $\mathcal{F}(B)$.

\subsection{The ontological fingerprint}

Each selected model $\Theta^{*(f)}$ carries a structured parameter set that
characterizes the physical behavior of the blanket under function $f$:

\begin{itemize}
  \item $A_{bb}^{(f)}$ and $\rho\!\left(A_{bb}^{(f)}\right)$ --- the spectral
        radius of the blanket's self-dynamics.
        A higher value means the surface retains memory of its state longer,
        consistent with thermal insulation (travel mug: $\rho \approx 0.52$)
        versus rapid heat exchange (espresso: $\rho \approx 0.47$).
  \item $A_{sb}^{(f)}$, $A_{bz}^{(f)}$ --- the coupling channels; their
        relative magnitudes encode whether the object is primarily
        \emph{receiving} energy from the interior or \emph{dissipating} it
        to the environment.
  \item $C_b^{(f)}$ --- the observation model for the surface, specifying
        which linear combination of the latent blanket state generates the
        observable measurements (temperature, curvature, flux).
  \item $\boldsymbol{\lambda}^{*(f)} = (\lambda_1^{*(f)}, \ldots,
        \lambda_K^{*(f)})$ --- the \emph{ontological fingerprint}: the vector
        of converged Lagrange multipliers encodes which physical properties
        the surface \emph{must} regulate in order to fulfill function $f$.
        For an espresso cup, the volume constraint $\lambda_{R4}$ and the
        thermal comfort constraint $\lambda_{R1}$ are both active; for a
        travel mug, $\lambda_{R1}$ dominates, encoding the primacy of
        thermal isolation.
\end{itemize}

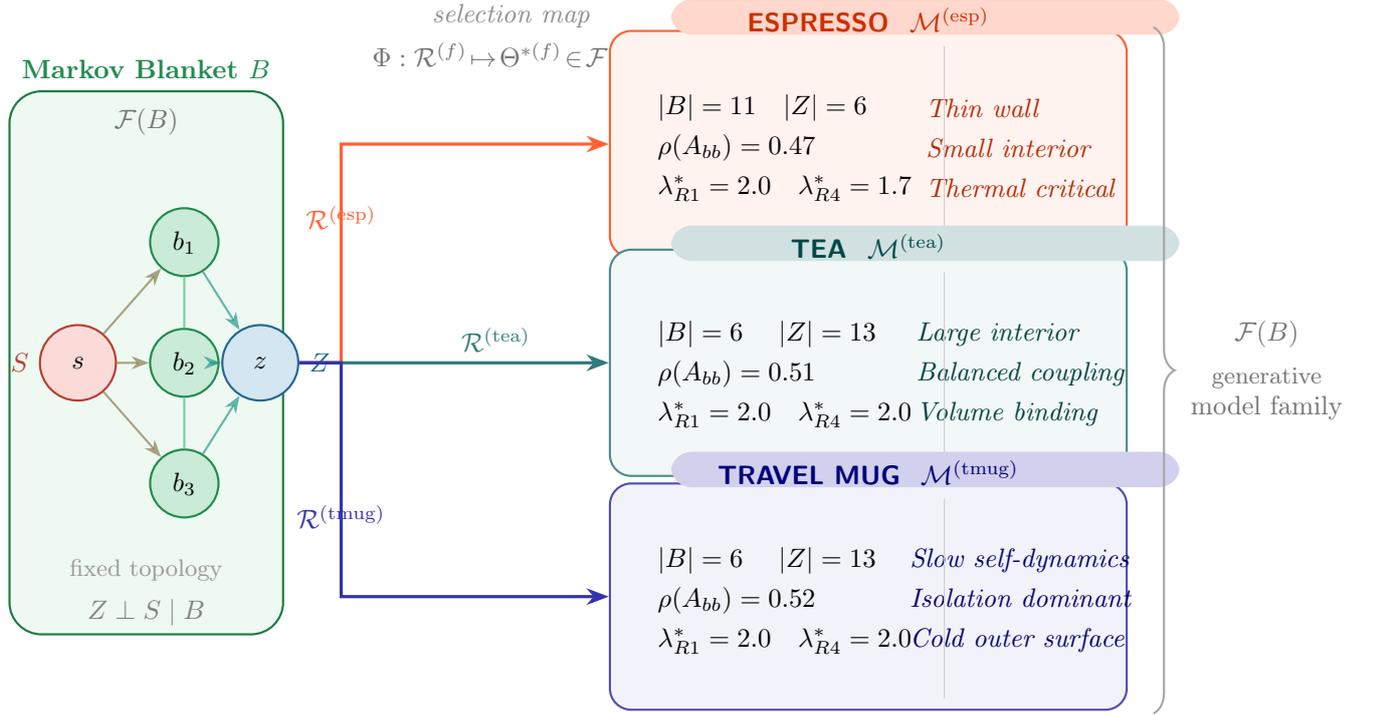
\begin{figure}[h]
\centering
\begin{tikzpicture}[
  every node/.style = {font=\normalsize},
  >=Stealth
]

\colorlet{espCol}{red!55!orange}
\colorlet{teaCol}{teal!70!black}
\colorlet{mugCol}{blue!60!black}


\node[
  draw=colB!70!black, fill=colB!8, thick,
  rounded corners=12pt,
  minimum width=3.6cm, minimum height=7.2cm
] (MBbox) at (0,0) {};

\node[font=\small\bfseries, text=colB!80!black]
  at (0, 3.9) {Markov Blanket $B$};

\node[circle, draw=colS!80!black, fill=colS!20, thick,
      minimum size=1.0cm, inner sep=0pt,
      label={[font=\small, text=colS!80!black]left:$S$}]
      (S) at (-0.9, 0) {\small$s$};

\node[circle, draw=colB!80!black, fill=colB!25, thick,
      minimum size=0.9cm, inner sep=0pt]
      (B1) at (0.5,  1.6) {\small$b_1$};
\node[circle, draw=colB!80!black, fill=colB!25, thick,
      minimum size=0.9cm, inner sep=0pt]
      (B2) at (0.5,  0.0) {\small$b_2$};
\node[circle, draw=colB!80!black, fill=colB!25, thick,
      minimum size=0.9cm, inner sep=0pt]
      (B3) at (0.5, -1.6) {\small$b_3$};

\node[circle, draw=colZ!80!black, fill=colZ!20, thick,
      minimum size=1.0cm, inner sep=0pt,
      label={[font=\small, text=colZ!80!black]right:$Z$}]
      (Z) at (1.5, 0) {\small$z$};

\foreach \b in {B1,B2,B3}{
  \draw[->, draw=colS!50!colB, thick, opacity=0.7] (S) -- (\b);
}
\foreach \b in {B1,B2,B3}{
  \draw[->, draw=colB!50!colZ, thick, opacity=0.7] (\b) -- (Z);
}
\draw[draw=colB!60, thick] (B1)--(B2)--(B3);

\node[font=\small\itshape, text=black!50]
  at (0, -3.3) {$Z \perp S \mid B$};
\node[font=\footnotesize, text=black!40, align=center]
  at (0, -2.75) {fixed topology};

\node[font=\small\bfseries, text=black!55, align=center]
  at (0, 3.2) {\small $\mathcal{F}(B)$};


\node[font=\small\itshape, text=black!45, align=center]
  at (4.8, 4.6) {selection map};
\node[font=\small, text=black!55, align=center]
  at (4.8, 4.05)
  {$\Phi:\mathcal{R}^{(f)}\!\mapsto\!\Theta^{*(f)}\!\in\!\mathcal{F}(B)$};


\node[
  draw=espCol!80, fill=espCol!6, thick,
  rounded corners=8pt,
  minimum width=6.8cm, minimum height=3.0cm
] (M1) at (9.5, 2.9) {};

\fill[espCol!20, rounded corners=8pt]
  (6.85+0.05, 2.9+1.5-0.05) rectangle (6.85+6.75, 2.9+1.5+0.42);
\node[font=\small\bfseries, text=espCol!80!black]
  at (9.5, 4.55) {\textsf{ESPRESSO}\ \ $\mathcal{M}^{(\mathrm{esp})}$};

\node[font=\small, align=left] at (8.4, 2.85) {%
  $|B| = 11$ \ \ $|Z| = 6$\\[3pt]
  $\rho(A_{bb}) = 0.47$\\[3pt]
  $\lambda^*_{R1} = 2.0$\quad$\lambda^*_{R4} = 1.7$};
\node[font=\small\itshape, text=espCol!70!black, align=left]
  at (11.5, 2.85)
  {Thin wall\\[3pt]Small interior\\[3pt]Thermal critical};

\draw[black!20, thin] (10.5, 1.55) -- (10.5, 4.2);

\node[
  draw=teaCol!70, fill=teaCol!5, thick,
  rounded corners=8pt,
  minimum width=6.8cm, minimum height=3.0cm
] (M2) at (9.5, -0) {};

\fill[teaCol!18, rounded corners=8pt]
  (6.85+0.05, -0.1+1.5-0.05) rectangle (6.85+6.75, -0.1+1.5+0.42);
\node[font=\small\bfseries, text=teaCol!80!black]
  at (9.5, 1.55) {\textsf{TEA}\ \ $\mathcal{M}^{(\mathrm{tea})}$};

\node[font=\small, align=left] at (8.4, -0.15) {%
  $|B| = 6$ \ \ \ $|Z| = 13$\\[3pt]
  $\rho(A_{bb}) = 0.51$\\[3pt]
  $\lambda^*_{R1} = 2.0$\quad$\lambda^*_{R4} = 2.0$};
\node[font=\small\itshape, text=teaCol!70!black, align=left]
  at (11.5, -0.15)
  {Large interior\\[3pt]Balanced coupling\\[3pt]Volume binding};

\draw[black!20, thin] (10.5, -1.45) -- (10.5, 1.2);

\node[
  draw=mugCol!70, fill=mugCol!5, thick,
  rounded corners=8pt,
  minimum width=6.8cm, minimum height=3.0cm
] (M3) at (9.5, -3.1) {};

\fill[mugCol!18, rounded corners=8pt]
  (6.85+0.05, -3.1+1.5-0.05) rectangle (6.85+6.75, -3.1+1.5+0.42);
\node[font=\small\bfseries, text=mugCol!80!black]
  at (9.5, -1.45) {\textsf{TRAVEL MUG}\ \ $\mathcal{M}^{(\mathrm{tmug})}$};

\node[font=\small, align=left] at (8.4, -3.15) {%
  $|B| = 6$ \ \ \ $|Z| = 13$\\[3pt]
  $\rho(A_{bb}) = 0.52$\\[3pt]
  $\lambda^*_{R1} = 2.0$\quad$\lambda^*_{R4} = 2.0$};
\node[font=\small\itshape, text=mugCol!70!black, align=left]
  at (11.5, -3.15)
  {Slow self-dynamics\\[3pt]Isolation dominant\\[3pt]Cold outer surface};

\draw[black!20, thin] (10.5, -4.45) -- (10.5, -1.8);


\draw[->, thick, draw=espCol!80, line width=1.2pt]
  (Z.east) -- ++(0.55,0) |- (M1.west)
  node[pos=0.28, above, font=\small\itshape, text=espCol!80]
  {$\mathcal{R}^{(\mathrm{esp})}$};

\draw[->, thick, draw=teaCol!80, line width=1.2pt]
  (Z.east) -- ++(1.1,0) --
  node[above, font=\small\itshape, text=teaCol!80]
  {$\mathcal{R}^{(\mathrm{tea})}$}
  (M2.west);

\draw[->, thick, draw=mugCol!80, line width=1.2pt]
  (Z.east) -- ++(0.55,0) |- (M3.west)
  node[pos=0.28, below, font=\small\itshape, text=mugCol!80]
  {$\mathcal{R}^{(\mathrm{tmug})}$};

\draw[decorate,
      decoration={brace, amplitude=8pt},
      thick, black!35]
  (13.25, 4.45) -- (13.25,-4.65)
  node[midway, right=10pt, font=\small, align=center, text=black!55]
  {$\mathcal{F}(B)$\\[4pt]\footnotesize generative\\model family};

\end{tikzpicture}

\caption{%
\textbf{A Markov blanket topology and two phenomena it supports.}
A fixed blanket topology $B$ (left) is compatible with a family
$\mathcal{F}(B)$ of generative models (right). The three models shown
illustrate \emph{two distinct phenomena}, not a single family.
\textsf{Tea} and \textsf{Travel Mug} share the same topology ($|B|{=}6$,
$|Z|{=}13$) and belong to the \emph{same family} $\mathcal{F}^{(1)}$:
they differ only in spectral radius $\rho(A_{bb})$ and Lagrange fingerprint
$\boldsymbol{\lambda}^*$ --- this is \textbf{intra-family navigation} (P1).
\textsf{Espresso}, by contrast, has a different topology ($|B|{=}11$,
$|Z|{=}6$) and belongs to a \emph{different family} $\mathcal{F}^{(2)}$:
moving from \textsf{Tea} to \textsf{Espresso} requires crossing the critical decision boundary
$\mathcal{H}$ in requirement space --- this is \textbf{family transition} (P2).
The selection map $\Phi: \mathcal{R}^{(f)} \mapsto \Theta^{*(f)}$ therefore
operates on two levels: within a family (navigation, changing $A_{bb}$, $C_b$,
$\boldsymbol{\lambda}^*$ at fixed topology) and across families (transition,
changing the blanket signature $\sigma$ itself).%
}
\label{fig:mb_family}
\end{figure}

Figure~\ref{fig:mb_family} illustrates the relationship between a single
Markov blanket topology and the family of generative models it spans across
three distinct functional requirement profiles.

\medskip

This picture has a precise design implication. When an engineer asks
``what is the optimal cup?'' the question is, strictly speaking, underdetermined
without specifying the intended function. C-DMBD makes this explicit: the
Markov blanket $B$ defines the \emph{interface type} --- the topological fact
that there is a surface mediating between interior and environment --- while the
generative model $\Theta^{*(f)} \in \mathcal{F}(B)$ defines the
\emph{functional mode} --- the specific physical behavior that surface must
exhibit to fulfill function $f$.

Two cups may share the same blanket topology (same nodes in $B$, same graph
connectivity) and yet be physically different objects in a principled sense:
they are different members of the same generative model family, selected by
different requirement profiles. Conversely, two cups with different blanket
topologies belong to different families altogether --- they are not just
differently tuned instances of the same object type, but objects of
categorically different ontological kinds.

Thus, the Markov blanket fixes the general structure of the product, while the generative model describes how this structure evolves over time. Now, a single structure can contain multiple possible models. The functional mode manifold is precisely the space of these possible models, but restricted to the parameters that can vary without changing the structure — that is, it specifies which models can function given that structure.

\section{Invariants, Variants, and the Geometry of a Generative Model Family}
\label{sec:geometry}

A blanket topology induces a family $\mathcal{F}(B)$ of generative models. To
make this precise we need three things: a definition of what is invariant across
the family, a characterization of the free parameters, and a formal criterion
for family membership.

\subsection{The blanket signature: invariant structure}

Fix a dataset $\mathbf{y}$ and run C-DMBD under two requirement profiles,
obtaining $(\boldsymbol{\omega}^{*(1)}, \Theta^{*(1)})$ and
$(\boldsymbol{\omega}^{*(2)}, \Theta^{*(2)})$. We say the two solutions belong
to the \emph{same family} if they share the same \emph{blanket signature}:

\begin{definition}[Blanket signature]
$\sigma(\boldsymbol{\omega}^*, \Theta^*) =
  \bigl(\boldsymbol{\omega}^*,\;\mathrm{supp}(A),\;\mathrm{block}(C)\bigr)$,
where $\mathrm{supp}(A)$ is the sparsity pattern of the dynamics matrix (with
MB-zeros $A_{sz}{=}\mathbf{0}$, $A_{zs}{=}\mathbf{0}$) and
$\mathrm{block}(C)$ is the block-diagonal structure of the emission matrices.
\end{definition}

Two solutions are in the same family iff $\sigma^{(1)} = \sigma^{(2)}$.
The signature encodes which channels carry information; it does not
specify how much flows through each.

\subsection{The functional mode manifold: what varies}

Within a fixed family, the free parameters are the numerical values of the
non-zero blocks:
\begin{equation}
  \boldsymbol{\theta} =
  \bigl(\mathrm{vec}(A_{bb}),\;
        \mathrm{vec}(A_{sb}),\;
        \mathrm{vec}(A_{bz}),\;
        \mathrm{vec}(C_b),\;
        \boldsymbol{\lambda}^*\bigr)
  \;\in\; \mathbb{R}^{d_\theta}.
  \label{eq:theta}
\end{equation}
The set of admissible $\boldsymbol{\theta}$ (spectrally stable $A$, positive
definite noise) forms a smooth manifold $\mathcal{M}_\mathcal{F}$, the
\emph{functional mode manifold}. The selection map
$\Phi: \mathcal{R} \to \mathcal{M}_\mathcal{F}$ sends each requirement profile
to a point on this manifold. Its Jacobian
$J_\Phi = \partial\boldsymbol{\theta}^*/\partial\mathcal{R}
\in \mathbb{R}^{d_\theta \times K}$
measures functional plasticity: how sensitively the mode responds to changes
in what is demanded of the interface.

Within $\mathcal{M}_\mathcal{F}$, variation decomposes into three layers.
\emph{Dynamical variation} ($A_{bb}$, $A_{sb}$, $A_{bz}$) governs coupling
strengths and temporal memory; $\rho(A_{bb})$ is the scalar summary.
\emph{Observational variation} ($C_b$) governs which physical signals the
blanket transmits. \emph{Ontological variation} ($\boldsymbol{\lambda}^*$)
records which requirements were binding at equilibrium, i.e.\ what the
surface had to actively work to satisfy.

\subsection{Formal criterion for family membership}

The gap $\Delta(\boldsymbol{\omega}^*)$ (defined in Section~\ref{sec:phenomena})
is the operational criterion: when the total Lagrange penalty differential
between $\boldsymbol{\omega}^*$ and the runner-up partition is smaller than
$\Delta$, the requirement profile cannot flip the partition and the family is
fixed. Different profiles then navigate $\mathcal{M}_\mathcal{F}$ without
changing $\sigma$. When the gap is near zero, small requirement changes can
trigger a transition to a different family --- the subject of Proposition~2.

\section{The Three Phenomena: Navigation, Transition, Disambiguation}
\label{sec:phenomena}

The three propositions below describe three distinct ways in which the requirements shape both the type of interface and the generative models compatible with it. They are \emph{conditional observations}, not
unconditional theorems. The joint optimization over
$(\boldsymbol{\omega}, \Theta, \boldsymbol{\lambda})$ is non-convex;
convergence to a local optimum is guaranteed by the VBEM structure, but
not to a global one. Each proposition holds under stated conditions that
can be verified empirically, and is falsifiable: an experiment that
satisfies the stated conditions but fails to exhibit the predicted
phenomenon would constitute a genuine counterexample.

A specific non-circularity argument applies to P2. The gap score
$\Delta(\boldsymbol{\omega}^*)$ is computed \emph{without} any requirements
(all $\lambda_k = 0$): it measures the data's inherent topological
information before any functional demand is applied. When $\Delta \approx 0$,
P2 \emph{predicts} that a continuous requirement scan will produce a
discontinuous topology change --- a prediction made from data alone, prior
to running any requirement-steered inference. The experiment then verifies
this prediction. The validation is therefore not circular: the predictive
instrument ($\Delta$) and the validated phenomenon (topology jump) are
computed in two separate, independent inference runs.

Let $p_\mathrm{data}(\boldsymbol{\omega})$ denote the data-driven posterior
over partitions (the ELBO with all $\lambda_k = 0$). The \emph{gap} at the
MAP partition $\boldsymbol{\omega}^*$ is
\[
  \Delta(\boldsymbol{\omega}^*) =
  \log p_\mathrm{data}(\boldsymbol{\omega}^*) -
  \max_{\boldsymbol{\omega} \neq \boldsymbol{\omega}^*}
  \log p_\mathrm{data}(\boldsymbol{\omega}).
\]

\subsection{Intra-family navigation}

\begin{proposition}[Navigation]
\label{prop:navigation}
Suppose $\Delta(\boldsymbol{\omega}^*) > 0$. Let
$\boldsymbol{\omega}' = \arg\max_{\boldsymbol{\omega} \neq \boldsymbol{\omega}^*}
\log p_\mathrm{data}(\boldsymbol{\omega})$ be the second-best partition.
Suppose two requirement profiles $\mathcal{R}^{(1)}$, $\mathcal{R}^{(2)}$
produce converged Lagrange vectors $\boldsymbol{\lambda}^{(f)}$ satisfying

\[
  \sum_{k=1}^K \lambda_k^{(f)}
  \bigl[v_k(\boldsymbol{\omega}^*, \Theta^{(f)}) -
         v_k(\boldsymbol{\omega}', \Theta^{(f)})\bigr]
  < \Delta(\boldsymbol{\omega}^*), \quad f = 1, 2.
\]

Then both profiles select the same interface type --- the same blanket
signature $\sigma^*$ --- but, through the gradient
$\nabla_\Theta\mathcal{L}_\mathrm{aug}$ (Eq.~\ref{eq:grad_theta}),
generically land on distinct functional modes:
$\boldsymbol{\theta}^{*(1)} \neq \boldsymbol{\theta}^{*(2)}$ on
$\mathcal{M}_\mathcal{F}$.
\end{proposition}

The condition now involves the \emph{total} penalty
differential across the two best partitions, not the maximum single term.
This is the correct quantity: a requirement profile whose total violation
advantage for $\boldsymbol{\omega}'$ over $\boldsymbol{\omega}^*$ does not
exceed the data gap will not flip the partition. Once the partition is
stabilized, the $\Theta$-gradient in Eq.~(\ref{eq:grad_theta}) drives the
M-step toward the point on $\mathcal{M}_\mathcal{F}$ that balances data fit
and requirement satisfaction for that profile.

\subsection{Family transition}

\begin{proposition}[Phase transition of interface type]
\label{prop:transition}
Suppose two partitions $\boldsymbol{\omega}^{(1)}$, $\boldsymbol{\omega}^{(2)}$
satisfy $\Delta(\boldsymbol{\omega}^{(1)}) \approx 0$. For a given
requirement profile $\boldsymbol{\lambda}$, define the \emph{objective
advantage} of $\boldsymbol{\omega}^{(1)}$ as
\[
  \mathcal{A}(\boldsymbol{\lambda}) =
  \bigl[\log p_\mathrm{data}(\boldsymbol{\omega}^{(1)}) -
         \log p_\mathrm{data}(\boldsymbol{\omega}^{(2)})\bigr]
  - \sum_k \lambda_k
    \bigl[v_k(\boldsymbol{\omega}^{(1)},\Theta^*(\boldsymbol{\omega}^{(1)},\boldsymbol{\lambda})) -
          v_k(\boldsymbol{\omega}^{(2)},\Theta^*(\boldsymbol{\omega}^{(2)},\boldsymbol{\lambda}))\bigr].
\]
C-DMBD selects $\boldsymbol{\omega}^{(1)}$ when $\mathcal{A} > 0$ and
$\boldsymbol{\omega}^{(2)}$ when $\mathcal{A} < 0$. The locus
\[
  \mathcal{H} = \{\boldsymbol{\lambda} : \mathcal{A}(\boldsymbol{\lambda}) = 0\}
\]
is the \emph{critical decision boundary} separating the two families in
$\boldsymbol{\lambda}$-space. Continuously deforming $\boldsymbol{\lambda}$
across $\mathcal{H}$ induces a discontinuous change of interface type.
\end{proposition}

Because $\Theta^*(\boldsymbol{\omega}, \boldsymbol{\lambda})$
depends on $\boldsymbol{\lambda}$ through the M-step equilibrium, the
objective advantage $\mathcal{A}(\boldsymbol{\lambda})$ is in general a
nonlinear function of $\boldsymbol{\lambda}$. Consequently, $\mathcal{H}$ is
in general a \emph{curved hypersurface}, not a hyperplane. It reduces to a
hyperplane only in the special case where the M-step parameters
$\Theta^*$ are insensitive to $\boldsymbol{\lambda}$ at fixed partition --- an
approximation that may hold locally but not globally. We use the term
``critical decision boundary'' throughout to avoid this overstatement.
The boundary cannot be computed analytically; it is estimated empirically
by scanning the requirement profile and detecting the topology change, as
in the P2 experiment.

\subsection{Ontological disambiguation}

Proposition~3 operates on two distinct levels that must be kept separate. 
Proposition~3a is a mathematical limit result: it establishes what the 
framework entails when data are completely uninformative. Proposition~3b 
is an operative result: it describes what the experiments in Section~11 
actually demonstrate. Conflating the two would overstate the empirical 
claim; separating them makes both more precise.

\begin{proposition3a}[Ontological disambiguation --- strong form]
Suppose $p_{\mathrm{data}}(\omega)$ is uniform over all partitions, 
i.e., $\Delta(\omega^*) = 0$ exactly. Then the augmented ELBO reduces 
to $-\sum_k \lambda_k v_k(\omega, \Theta)$, and C-DMBD converges to
\begin{equation}\label{eq:p3a}
(\omega^*, \Theta^*) = \arg\min_{\omega,\,\Theta} 
\sum_k \lambda^*_k(\omega,\Theta)\cdot v_k(\omega,\Theta),
\end{equation}
where $\lambda^*(\omega,\Theta)$ are the Lagrange multipliers converged 
at $(\omega,\Theta)$. In this limit the interface type is determined 
entirely by the requirements: C-DMBD \emph{constructs} the type rather 
than \emph{discovers} it. Requirements have direct ontological force.
\end{proposition3a}

Proposition~3a is a theoretical horizon, not an observable regime. The 
flat-data limit $\Delta = 0$ exactly is a mathematical idealization: 
real physical data always carry some topological information, however 
weak. The strong form establishes that the logical structure of the 
framework admits requirement-determined types; it does not claim that 
physical experiments will reach this limit.

\begin{proposition3b}[Ontological disambiguation --- operative form]
Suppose $\Delta(\omega^*) \approx 0$ but $\Delta > 0$. Then the data 
provide weak but non-zero topological information. In this regime, the 
requirement profile is the dominant force shaping the interface type: 
small changes in $\{R_k\}$ produce changes in $\omega^*$ that the data 
alone, at $\lambda = 0$, would not produce. C-DMBD resolves the 
topological ambiguity that the data cannot settle, and the resolved 
type is requirement-determined to first order.
\end{proposition3b}

Proposition~3b is the regime demonstrated in Section~11: at gap 
$\Delta \approx 0.17$, Espresso and Travel Mug produce distinct 
topologies ($|B|=10$ vs $|B|=8$) under the same physical data. The 
data provide insufficient discrimination; the requirement profile 
supplies the remainder. This is already a non-trivial result: 
functional specifications can determine what kind of thing an object 
is, not merely how that thing performs --- even when the data are not 
entirely silent. Proposition~3a is the logical completion of 
Proposition~3b: it shows what the operative phenomenon converges to 
as $\Delta \to 0$. The two forms are not independent claims but two 
readings of the same underlying structure, one exact and one 
approximate, each illuminating the other.

\subsection{The certificate of functional effort}

All three propositions share a common object: the converged Lagrange vector
$\boldsymbol{\lambda}^*$. We give it a name and a precise formal interpretation.

\begin{definition}[Certificate of functional effort]
The \emph{certificate of functional effort} of a C-DMBD solution is the
converged Lagrange vector $\boldsymbol{\lambda}^* \in \mathbb{R}^K_{\geq 0}$.
$\lambda_k^* = 0$ certifies that $\mathcal{R}_k$ is satisfied at the joint
fixed point $(\boldsymbol{\omega}^*, \Theta^*, \boldsymbol{\lambda}^*)$
without inferential pressure: the topology and the data-preferred
parameterization satisfy it freely. $\lambda_k^* > 0$ certifies that
$\mathcal{R}_k$ remains violated at magnitude $v_k(\boldsymbol{\omega}^*,
\Theta^*)$ even after the M-step has optimized $\Theta$ to minimize the
violation through the gradient $\nabla_\Theta \mathcal{L}_\mathrm{aug}$.
\end{definition}

The formal distinction from classical Lagrange multipliers is the following.
In constrained parametric optimization, $\mu_j^* = -\partial f(p^*)/\partial
\epsilon_j$: the multiplier measures how much the optimal \emph{cost} would
improve if constraint $g_j$ were relaxed, evaluated at the cost optimum $p^*$.
The C-DMBD certificate $\lambda_k^*$ is not a cost sensitivity. It is the
violation level $v_k(\boldsymbol{\omega}^*, \Theta^*)$ at the fixed point of
the dual ascent update, i.e.\ the violation that persists after the
\emph{best achievable parameterization of this topology}, given the data, has
already been found. Formally, at convergence:
\[
  \lambda_k^* = \mathrm{clip}\!\left(
    \lambda_k^* + \alpha_\lambda \cdot v_k(\boldsymbol{\omega}^*, \Theta^*),
    0, \lambda_\mathrm{max}
  \right),
\]
which at the fixed point implies $v_k(\boldsymbol{\omega}^*, \Theta^*) = 0$
when $\lambda_k^* < \lambda_\mathrm{max}$, or $v_k > 0$ when the multiplier
is saturated. A saturated $\lambda_k^*$ means the topology cannot satisfy
$\mathcal{R}_k$ at any feasible parameterization within the data-constrained
model --- not that the cost is sensitive to $\mathcal{R}_k$. An unsaturated
$\lambda_k^* > 0$ (rare in practice) means the requirement is satisfied at
the fixed point but required non-zero pressure during convergence.

In this sense the certificate records \emph{parametric resistance at the
type equilibrium}: the degree to which the best achievable member of the
generative model family $\mathcal{F}(B_\mathrm{cup})$, given the observed
data, still fails to realize the functional requirement. It is therefore
informative \emph{comparatively} --- the difference in certificate values
across cup styles, or across requirement dimensions within the same style,
is the primary quantity of interest. Absolute magnitudes are confounded by
the ceiling $\lambda_\mathrm{max}$ and the step size $\alpha_\lambda$, as
noted in the experimental results.

\section{Beyond Requirements: The Design Loop}
\label{sec:b2b}

Before tracing the design loop, a clarification on levels is 
necessary. C-DMBD as defined in Section~5 takes two inputs: 
observational data $y$ and a requirement profile $\{R_k\}$. In this section we introduce another element: the designer's generative model of the user, i.e., the set of expectations the designer has about the user's behavior. The 
user model $\tilde{p}_{\mathrm{user}}$ is not an input to the 
algorithm. The designer 
uses her model of the user to interpret the requirements, which are 
the sole channel through which expectations about the user enter 
C-DMBD. However, requirements are not generated by the designer's model. Furthermore, the designer must respect the constraints imposed by the C-DMBD and the family of generative models identified by the algorithm. This means that the designer must adapt their user model to the range of models defined by the Markov blanket.
 
\subsection{The design loop}
 
The framework developed so far treats the environment $S$ as a monolithic
block, collapsing hand, air, and table (in the case of a cup) into a single undifferentiated
partition. To go further, we need to distinguish precisely between what the
designer knows, what the cup communicates, and what the user can infer.
The result is a closed loop that we
now trace step by step.
 
\medskip
\noindent\textbf{Step 1: the designer infers the cup's blanket.}
The designer observes physical data $\mathbf{y}$ from the cup --- thermal
profiles, geometric measurements, flux fields --- and applies C-DMBD to infer
a Markov blanket $B_\text{cup}$ and the associated family of generative models
$\mathcal{F}(B_\text{cup})$. Each member $\Theta \in \mathcal{F}(B_\text{cup})$
is a distinct parameterization of the same blanket topology: a different set
of dynamics $A_{bb}$, emissions $C_b$, and functional effort certificate
$\boldsymbol{\lambda}^*$. In a nutshell, the Markov blanket identifies a family of admissible models. It's then up to the designer to identify the generative model of the cup best suited to her needs. It's worth clarifying: the Markov blanket opens up a range of possible models of the cup's behavior, but it's up to the designer to choose the one that best fits her preferences and goals. That model obviously doesn't belong to the cup — it belongs to the designer.
The cup is, in the FEP sense, an inert object. The design problem is: \textit{how can the user reconstruct the generative model of the cup based solely on the Markov blanket she can see?}
 
\medskip
\noindent\textbf{Step 2: the designer selects a member of the family.}
The designer also holds a second generative model: her model of the
cup--user relationship. This model is far more general and concerns user behavior: not just how the user physically interacts with the cup, but also the preferences that a given user (e.g., young people) may have for a certain aesthetic or lifestyle, or a particular set of cultural habits or features (e.g., fashion, movies).\footnote{Obviously, at the level of modeling, one can also speak of a single generative model with multiple layers. But there are other solutions as well. The intent here is simply to distinguish between the designer's expectations about the cup's behavior and the designer's expectations about the user's behavior. The designer must find a balance between these two sets of expectations. The point is that C-DMBD sets limits, constraints within which the designer can operate.}
 
\medskip
\noindent\textbf{Step 3: the user infers the designer's model from the blanket.}
The user has no access to the designer's models. The only channel available
to her is $B_\text{cup}$ --- the physical surface, with its stationary
distribution of temperature, pressure, and curvature
$p_\text{cup}(y_b \mid \Theta^{*(f)}) =
\mathcal{N}(\hat{\boldsymbol{\mu}}_b^{(f)}, \hat{\Sigma}_b^{(f)})$.
In the FEP framework, the user's hand is an autonomous
agent that minimizes variational free energy through active inference.
Her blanket $B_\text{user}$ has a Fristonian structure with three components:
\emph{sensory states} $o \subset B_\text{user}$ (skin temperature,
mechanoreceptor discharge, proprioception --- what the hand receives from
the cup without acting on it), \emph{active states} $a \subset B_\text{user}$
(grip force, finger repositioning, wrist adjustment --- what the hand does
to the cup, driven by internal predictions), and \emph{internal states}
$Z_\text{user}$ (the user's generative model of expected contact: priors
over temperature, pressure, curvature, updated by sensory states and
expressed through active states). Through an active inference loop, the user reconstructs --- from the
surface of the cup alone --- the designer's generative models of the cup and the user herself. She learns
what kind of object the cup is, what it expects of her hand and mouth (i.e., the coordination between hand and mouth), how it should
be gripped and used. The Markov blanket of the cup is therefore the sole
communication channel between the designer's intentions and the user's
inference.
 
\medskip
\noindent\textbf{Step 4: the circle closes.}
The user's actual generative model $\tilde{p}_\text{user}(y_h)$ --- the
prior her hand brings to the contact --- can itself be inferred from
physiological data (grip force time series, skin temperature, EMG) using
DMBD applied to the user system, recovering $B_\text{user}$, its
active/sensory partition, and the corresponding prior.
This closes the loop: the designer's prior model of the cup and the user can be compared with the
user's actual generative model discovered from data. The gap between them
is the measure of how well the designer understood the user before building
the object. A design is optimal when this gap is zero: when the surface the
designer built communicates exactly what the user's active inference expects
to find. In active inference terms, this process can also be described as synchronization between generative models \cite{FristonFrith2015}.

\medskip

In short, the designer's path runs from form to function: given a Markov blanket, she selects the generative model that encodes the intended use. The user's path runs in reverse: given only the form — the physical surface and its stationary statistics — she must reconstruct the function the designer had in mind.
  
\subsection{Optimal design as expected free energy minimization}
 
Under the active inference framework, the user minimizes her
\emph{expected free energy} $\mathcal{G}$ with respect to future contact
observations \cite{friston2019}:
 
\begin{equation}
  \mathcal{G}(\Theta^{*(f)}) =
  D_\mathrm{KL}\!\bigl(
    \tilde{p}_\text{user}(y_h) \;\|\;
    p_\text{cup}(y_b \mid \Theta^{*(f)})
  \bigr)
  -
  \mathbb{H}\!\bigl[\tilde{p}_\text{user}(y_h)\bigr].
  \label{eq:expected_free_energy}
\end{equation}
 
The first term is the KL divergence between what the user expects and what
the cup delivers --- the predicted surprise upon contact, the design gap.
The second term is the user's irreducible entropy, independent of the cup.
The optimal design is therefore:
 
\begin{definition}[Optimal design]
\begin{equation}
  \Theta^{**} = \arg\min_{\Theta \in \mathcal{F}(B_\text{cup})}
  D_\mathrm{KL}\!\bigl(
    \tilde{p}_\text{user}(y_h) \;\|\;
    p_\text{cup}(y_b \mid \Theta)
  \bigr).
  \label{eq:optimal_design}
\end{equation}
\end{definition}
 
\noindent The optimal cup minimises the effort required of
the user's hand: the surface communicates the designer's model of the
cup--user relationship so clearly that the user's internal model needs no
revision and her active states need no compensation. Design succeeds when
the surface speaks the user's language.
 
Note the asymmetry: the cup does not minimise surprise; the user does.
The designer's task is to select, from among the physically realisable
members of $\mathcal{F}(B_\text{cup})$, the one that minimises the user's
expected free energy. This is not requirement satisfaction. It is model
alignment.

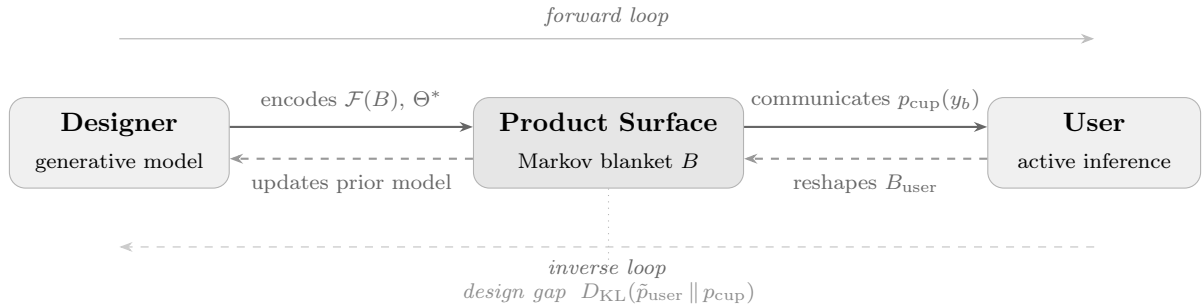
\begin{figure}[ht]
\centering
\begin{tikzpicture}[
    node distance = 3.2cm,
    box/.style = {
        rectangle, rounded corners=6pt,
        minimum width=2.8cm, minimum height=1.2cm,
        text centered, font=\small,
        draw=black!30, line width=0.4pt
    },
    designer/.style = {box, fill=black!6},
    object/.style   = {box, fill=black!10},
    user/.style     = {box, fill=black!6},
    fwd/.style  = {-{Stealth[length=5pt]},
                   line width=0.8pt, black!60},
    inv/.style  = {-{Stealth[length=5pt]},
                   line width=0.8pt, dashed, black!40},
    lbl/.style  = {font=\scriptsize, black!60},
    sublbl/.style = {font=\scriptsize\itshape, black!45}
]

\node[designer] (D) {%
    \begin{tabular}{c}
        \textbf{Designer}\\[2pt]
        \scriptsize generative model
    \end{tabular}};

\node[object, right=of D] (O) {%
    \begin{tabular}{c}
        \textbf{Product Surface}\\[2pt]
        \scriptsize Markov blanket $B$
    \end{tabular}};

\node[user, right=of O] (U) {%
    \begin{tabular}{c}
        \textbf{User}\\[2pt]
        \scriptsize active inference
    \end{tabular}};

\draw[fwd] ([yshift=6pt]D.east)
    -- node[lbl, above=2pt]{encodes $\mathcal{F}(B),\,\Theta^*$}
    ([yshift=6pt]O.west);

\draw[fwd] ([yshift=6pt]O.east)
    -- node[lbl, above=2pt]{communicates $p_{\mathrm{cup}}(y_b)$}
    ([yshift=6pt]U.west);

\draw[inv] ([yshift=-6pt]U.west)
    -- node[lbl, below=2pt]{reshapes $B_{\mathrm{user}}$}
    ([yshift=-6pt]O.east);

\draw[inv] ([yshift=-6pt]O.west)
    -- node[lbl, below=2pt]{updates prior model}
    ([yshift=-6pt]D.east);

\node[sublbl, below=1.1cm of O] (gap) {%
    design gap $\;D_{\mathrm{KL}}(\tilde{p}_{\mathrm{user}} \,\|\,
    p_{\mathrm{cup}})$};
\draw[dotted, black!30, line width=0.4pt]
    (O.south) -- (gap.north);

\draw[fwd, black!25, line width=0.5pt]
    ([yshift=22pt]D.north) --
    node[lbl, above=1pt]
        {\itshape forward loop}
    ([yshift=22pt]U.north);

\draw[inv, black!20, line width=0.5pt]
    ([yshift=-22pt]U.south) --
    node[lbl, below=1pt]
        {\itshape inverse loop}
    ([yshift=-22pt]D.south);

\end{tikzpicture}

\caption{\textbf{The design loop}. The forward loop (solid arrows) runs from
the designer's generative model through the product's Markov blanket
$B$ to the user's generative model: the designer encodes a functional
mode $\Theta^* \in \mathcal{F}(B)$ into the surface; the user
reconstructs it from the contact distribution
$p_{\mathrm{cup}}(y_b)$. The design gap
$D_{\mathrm{KL}}(\tilde{p}_{\mathrm{user}} \| p_{\mathrm{cup}})$
measures the residual mismatch. The inverse loop (dashed arrows)
runs in the opposite direction: the user's updated generative model
$\tilde{p}^{(t)}_{\mathrm{user}}$, inferred from physiological data,
flows back to the designer and steers a
new C-DMBD inference.}
\label{fig:design_loop}
\end{figure}
 
\subsection{Blanket compatibility and the extended objective}
 
When both distributions are linear-Gaussian, Eq.~(\ref{eq:optimal_design})
is computable in closed form. For a symmetric scalar summary of mutual fit:
 
\begin{equation}
  \mathcal{C}(B_\text{cup}, B_\text{user}) =
  -\tfrac{1}{2}\Bigl[
    D_\mathrm{KL}\bigl(p_\text{cup} \;\|\; \tilde{p}_\text{user}\bigr)
    +
    D_\mathrm{KL}\bigl(\tilde{p}_\text{user} \;\|\; p_\text{cup}\bigr)
  \Bigr].
  \label{eq:compatibility}
\end{equation}
 
The full augmented objective combining data fit, requirement satisfaction,
and user compatibility is:
 
\begin{equation}
  \mathcal{L}_\text{b2b} =
  \mathcal{L}_\text{ELBO}
  - \sum_k \lambda_k\, v_k(\boldsymbol{\omega}, \Theta)
  + \gamma\, \mathcal{C}(B_\text{cup}(\boldsymbol{\omega},\Theta),\;
                         B_\text{user}),
  \label{eq:elbo_b2b}
\end{equation}
 
where $\gamma > 0$ and $B_\text{user}$ is held fixed as a datum inferred
from physiological observations.
 
Two users with different priors $\tilde{p}_\text{user}^A \neq
\tilde{p}_\text{user}^B$ may require not just different parameter settings
but cups belonging to different families altogether: different interface
types, because their active inference systems carve the contact space
differently. The design loop runs separately for each user, and the gap
at Step 4 tells the designer how much her prior model of the user was wrong.
 
Equation~(\ref{eq:compatibility}) uses instantaneous stationary distributions.
Comparing power spectral densities $S_b^\text{cup}(e^{i\xi})$ and
$S_h^\text{user}(e^{i\xi})$ would additionally capture timescale mismatches
between the cup's thermal dynamics and the user's somatosensory expectations.
This is a natural next step not implemented in the current framework.

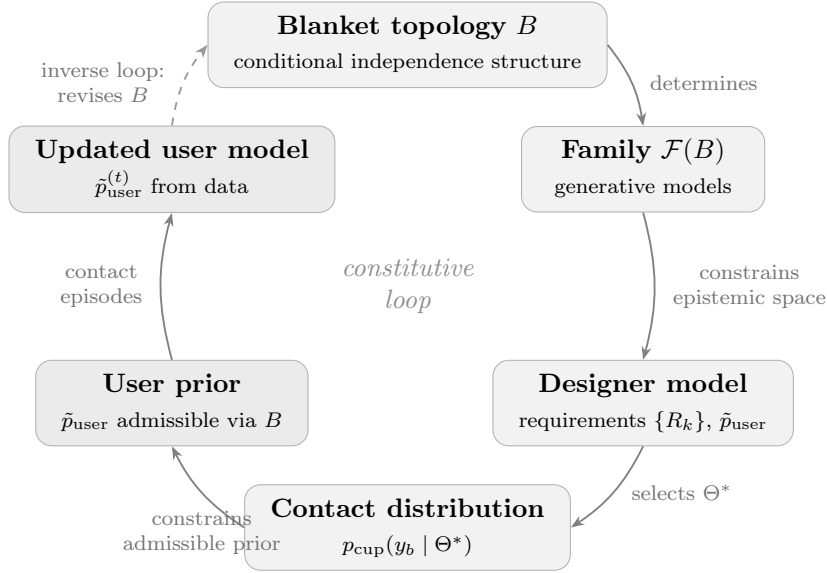
\begin{figure}[ht]
\centering
\begin{tikzpicture}[
    node distance = 0cm,
    box/.style = {
        rectangle, rounded corners=6pt,
        minimum width=3.2cm, minimum height=1.0cm,
        text centered, font=\small,
        draw=black!30, line width=0.4pt,
        fill=black!5
    },
    boxu/.style = {box, fill=black!8},
    arr/.style  = {-{Stealth[length=5pt]},
                   line width=0.7pt, black!50},
    arrinv/.style = {-{Stealth[length=5pt]},
                     line width=0.7pt, dashed, black!40},
    lbl/.style  = {font=\scriptsize, black!55,
                   align=center}
]


\node[box] (B)   at ( 0.00,  3.20) {%
    \begin{tabular}{c}
        \textbf{Blanket topology} $B$\\[1pt]
        \scriptsize conditional independence structure
    \end{tabular}};

\node[box] (FB)  at ( 3.10,  1.55) {%
    \begin{tabular}{c}
        \textbf{Family} $\mathcal{F}(B)$\\[1pt]
        \scriptsize generative models
    \end{tabular}};

\node[box] (DM)  at ( 3.10, -1.55) {%
    \begin{tabular}{c}
        \textbf{Designer model}\\[1pt]
        \scriptsize requirements $\{R_k\}$,
        $\tilde{p}_{\mathrm{user}}$
    \end{tabular}};

\node[box] (PC)  at ( 0.00, -3.20) {%
    \begin{tabular}{c}
        \textbf{Contact distribution}\\[1pt]
        \scriptsize $p_{\mathrm{cup}}(y_b \mid \Theta^*)$
    \end{tabular}};

\node[boxu] (UP)  at (-3.10, -1.55) {%
    \begin{tabular}{c}
        \textbf{User prior}\\[1pt]
        \scriptsize $\tilde{p}_{\mathrm{user}}$
        admissible via $B$
    \end{tabular}};

\node[boxu] (UU)  at (-3.10,  1.55) {%
    \begin{tabular}{c}
        \textbf{Updated user model}\\[1pt]
        \scriptsize $\tilde{p}^{(t)}_{\mathrm{user}}$
        from data
    \end{tabular}};


\draw[arr]
    (B.east)  to[bend left=15]
    node[lbl, right=3pt]{determines}
    (FB.north);

\draw[arr]
    (FB.south) to[bend left=15]
    node[lbl, right=3pt]{constrains\\epistemic space}
    (DM.north);

\draw[arr]
    (DM.south) to[bend left=15]
    node[lbl, right=3pt]{selects $\Theta^*$}
    (PC.east);

\draw[arr]
    (PC.west) to[bend left=15]
    node[lbl, below=3pt]{constrains\\admissible prior}
    (UP.south);

\draw[arr]
    (UP.north) to[bend left=15]
    node[lbl, left=3pt]{contact\\episodes}
    (UU.south);


\draw[arrinv]
    (UU.north) to[bend left=15]
    node[lbl, left=3pt]{inverse loop:\\revises $B$}
    (B.west);

\node[font=\small\itshape, black!45, align=center] at (0,0) {%
    constitutive\\loop};

\end{tikzpicture}

\caption{\textbf{The complete and constitutive design loop}. The blanket topology $B$ determines
the family $\mathcal{F}(B)$ of generative models, which constitutes
the designer's epistemic space --- the set of user--object
relationships the designer can represent. This space shapes the
admissible requirements $\{R_k\}$ and user prior
$\tilde{p}_{\mathrm{user}}$, which select a functional mode
$\Theta^* \in \mathcal{F}(B)$ and thus a contact distribution
$p_{\mathrm{cup}}(y_b \mid \Theta^*)$. The contact distribution in
turn constrains which user priors are admissible --- feeding back to
$B$. The inverse loop (dashed) closes when physiological contact
data update $\tilde{p}^{(t)}_{\mathrm{user}}$, which either revises
$\Theta^*$ within the current family or triggers a family transition
to a new blanket topology $B'$.}
\label{fig:constitutive_loop}
\end{figure}
\section{Comparison with Classical Parametric Design and Related Methods}

The methods reviewed in this section are mature and powerful. Rather than 
listing what each cannot do, we organize the comparison around a single 
question: under what conditions does C-DMBD reduce to an existing method? 
This structure reveals that C-DMBD recovers each framework as a limit case, 
while handling a class of problems that none of them can pose. The limits 
are not rhetorical devices --- they are identifiable regimes with 
verifiable conditions, and they delimit precisely where the existing 
methods are fully adequate and where they are not.

\subsection{Limit 1: large gap, weak requirements --- recovery of DMBD}

When the data gap $\Delta(\omega^*)$ is large and all Lagrange multipliers 
are initialized to zero ($\lambda = 0$), the augmented ELBO reduces to the 
standard ELBO and C-DMBD converges to the same solution as unsupervised 
DMBD \cite{beck2025}. Requirements play no role; the interface type is 
determined entirely by the data. This is the regime DMBD was designed for, 
and C-DMBD offers no advantage over it. The recovery is exact, not 
approximate: the two algorithms are identical in this limit. This confirms 
that C-DMBD is a proper extension of DMBD, not a replacement --- the 
requirement machinery is simply inert when there are no requirements to 
satisfy.

\subsection{Limit 2: infinite gap, fixed type --- proximity to parametric 
optimization}

When $\Delta(\omega^*) \to \infty$, no requirement profile can shift the 
partition: the data determine the interface type absolutely. C-DMBD then 
navigates the functional mode manifold $\mathcal{M}_F$ at fixed topology, 
optimizing $\Theta$ to satisfy the requirement profile. This is the regime 
most closely analogous to classical parametric optimization 
\cite{papalambros2017, nocedal2006}:
\begin{equation}
p^* = \arg\min_{p \in \mathcal{P}}\; f(p) 
\quad \text{subject to} \quad g_j(p) \leq 0,\; j = 1,\dots,J,
\end{equation}
which finds optimal parameters within a fixed, pre-specified type.

The analogy holds at the level of structure --- both methods optimize 
parameters at fixed topology --- but breaks down in three respects that 
persist even in this limit. First, parametric optimization presupposes 
a physics model; C-DMBD infers it from data. Second, the classical 
Lagrange multiplier $\mu^*_j$ is a cost sensitivity: it measures how much 
the optimal value of $f$ would change if constraint $g_j$ were relaxed by 
$\epsilon$ at the cost optimum. The C-DMBD certificate $\lambda^*_k$ is 
categorically different: it measures violation persistence at the type 
equilibrium, i.e., the degree to which the best achievable parameterization 
of this topology, given the data, still fails to realize requirement $R_k$. 
A high $\lambda^*_k$ indicates structural misalignment between the inferred 
interface and the requirement --- not proximity to a constraint boundary. 
No sensitivity coefficient in parametric optimization can express this, 
because parametric optimization has no notion of type and therefore no 
notion of type resistance. Third, the conditional independence structure 
$Z \perp S \mid B$ --- the causal fact that constitutes the interface as 
a boundary rather than a material region --- is not representable in a 
parametric design space $\mathcal{P} \subset \mathbb{R}^m$.

\subsection{Limit 3: known physics, continuous density --- proximity to 
topology optimization}

Topology optimization --- in particular SIMP \cite{bendsoe2003} and 
level-set methods --- produces topologies as output rather than 
presupposition, and is therefore a more demanding baseline. In SIMP, a 
density field $\rho(x) \in [0,1]$ is optimized over a fixed material 
domain under a known FEM physics model, penalizing intermediate densities 
to recover a binary (solid/void) topology \cite{ZienkiewiczTaylorZhu2013}. The limit case in which SIMP 
and C-DMBD most closely overlap is a binary SIMP solution on a system 
whose physics happen to generate a strong blanket signal in the data: in 
this regime, both methods locate the interface in approximately the same 
region, though by different procedures and for different reasons.

Outside this limit, three distinctions remain. First, topology 
optimization optimizes against a known, pre-specified physics model; 
C-DMBD infers the model jointly with the partition from observational 
data. The inference is Bayesian --- the partition is a latent variable 
with a posterior distribution, not a design variable with a cost function. 
Second, a SIMP topology is a spatial distribution of material density; a 
C-DMBD type is a conditional independence structure that determines which 
causal channels exist between interior and exterior. These are not 
competing representations of the same object: they are answers to 
different questions. Third, when $\Delta \approx 0$ and the data are 
consistent with multiple interface types, topology optimization has no 
mechanism to detect or resolve the ambiguity --- it optimizes a single 
objective and converges to one solution. C-DMBD detects the ambiguity 
through the gap score and resolves it through the requirement profile 
(Proposition~3b). This capability does not exist in any topology 
optimization framework, because those frameworks have no notion of 
topological ambiguity as a property of the data.

\subsection{Limit 4: user as boundary condition --- proximity to human 
factors engineering}

Decades of human factors engineering and computational ergonomics have 
produced rich and validated models of the user: anthropometric databases, 
biomechanical simulations, posture prediction tools, and comfort threshold 
models \cite{pheasant2006}. The present framework does not replace them, 
and in many practical contexts they remain the appropriate choice.

The limit case in which the two approaches converge is the following: 
replace C-DMBD's blanket-to-blanket compatibility term 
(Eq.~\ref{eq:compatibility}) with a finite set of threshold constraints on the 
blanket's stationary statistics, and replace the inferred user model with 
a designer-specified anthropometric profile. In this limit, C-DMBD 
reduces to requirement-steered inference with ergonomic constraints --- 
a formulation that is structurally close to human factors engineering 
applied to surface design.

Outside this limit, the distinction is not quantitative but structural. 
In human factors engineering the user enters the design problem as a 
specification: a set of thresholds the object must satisfy. The user is 
a boundary condition, fixed before the problem is written down. In 
C-DMBD with blanket-to-blanket compatibility, the user is a physical 
system with its own Markov blanket $B_{\mathrm{user}}$, generative model 
$(A_{hh}, C_h)$, and stationary distribution $\tilde{p}_{\mathrm{user}}
(y_h)$ inferred from physiological data. The design target is not 
threshold satisfaction but distributional convergence: 
$\mathcal{C}(B_{\mathrm{cup}}, B_{\mathrm{user}}) \to 0$. The user is 
not specified --- she is discovered. This difference matters practically: 
the design loop (Section~\ref{sec:b2b}) closes only when the 
designer's prior model of the user can be compared with the user's actual 
generative model inferred from data. Human factors engineering has no 
equivalent closing condition, because it has no notion of the user's 
interior states as a dynamical system with a boundary that the designer's 
model could be wrong about.


\section{Implementation and Demonstration}
\label{sec:implementation}


\subsection{Simulation Setup}
\label{subsec:setup}

The theoretical claims of this paper turn on
properties of C-DMBD that are difficult to isolate in hardware experiments:
one needs to vary a single requirement while holding the physical substrate
constant, or make the data spatially uninformative by construction.
We therefore evaluate the theory through a controlled synthetic simulator
in which both the ground-truth Markov-blanket structure and the data-generating
parameters are known exactly, making it possible to verify that the algorithm's
behavior matches the predicted phenomena precisely.\footnote{%
  Full source code is available at \url{https://github.com/DesignAInf/DESIGN-CUPS-BLANKET/tree/main/cdmbd_sim}.
  All experiments are reproducible via \texttt{pip install cdmbd-sim \&\& cdmbd-run}. These experiments are intended as a computational illustration of the
  theory's internal consistency, not as an empirical validation.
}

Each cup is discretized into $N = 30$ nodes along the vertical axis.
Each node $i$ emits a $D = 7$-dimensional observation vector
\begin{equation}
  \mathbf{y}_i(t)
  =
  \bigl[
    T_\mathrm{inner},\;
    T_\mathrm{outer},\;
    \nabla T,\;
    q,\;
    r,\;
    \kappa,\;
    w
  \bigr]^\top \!\in \mathbb{R}^7
  \label{eq:obs_channels}
\end{equation}
at each of $T = 50$ time steps, where $T_\mathrm{inner}$, $T_\mathrm{outer}$
are surface temperature proxies (\si{\degreeCelsius}), $\nabla T$ the
cross-wall gradient (\si{\degreeCelsius\per\metre}), $q$ the heat flux
(\si{\watt\per\metre\squared}), $r$ the node radius (\si{\metre}),
$\kappa = 1/r$ the curvature, and $w$ the wall thickness (\si{\metre}).
Observations are generated by a linear-Gaussian state-space model whose
true dynamics matrix $\mathbf{A}_\mathrm{true}$ has the correct Markov-blanket
zero structure ($\mathbf{A}_{zs} = \mathbf{0}$, $\mathbf{A}_{sz} = \mathbf{0}$).
Three cup styles are parameterized as in Table~\ref{tab:styles}; the
\emph{double-wall} variant reduces $T_\mathrm{outer}$ to approximately
$31\,\si{\degreeCelsius}$, matching the insulation level of commercially
available vacuum cups.

\begin{table}[ht]
\centering
\caption{%
  Simulated cup styles.
  $\rho_\mathrm{true}$: spectral radius of the ground-truth blanket dynamics matrix.
  Superscripts \textnormal{sw}/\textnormal{dw} denote single-/double-wall construction.
}
\label{tab:styles}
\small
\begin{tabular}{lcccc}
\toprule
Style       & Vol.\ (ml)
            & $T_\mathrm{inner}$ (\si{\degreeCelsius})
            & $T_\mathrm{outer}^{\mathrm{sw/dw}}$ (\si{\degreeCelsius})
            & $\rho_\mathrm{true}$ \\
\midrule
Espresso   & 60  & 90 & 86.0 / 30.9 & 0.47 \\
Tea        & 250 & 85 & 83.0 / 31.0 & 0.51 \\
Travel mug & 350 & 80 & 86.0 / 30.9 & 0.52 \\
\bottomrule
\end{tabular}
\end{table}

The C-DMBD loop is implemented with four choices:

\smallskip\noindent
\emph{E-step.}
Node $i$ is assigned to the role $\ell^* \in \{S, B, Z\}$ maximizing
\begin{equation}
  \mathrm{score}(i,\, \ell)
  \;=\;
  -\frac{\|\bar{\mathbf{y}}_i - \boldsymbol{\mu}_\ell\|^2}{2\sigma^2}
  \;-\;
  \|\boldsymbol{\lambda}\|_1
  \cdot
  \max\!\bigl(0,\;\delta_B(i)\bigr)\cdot\mathbf{1}[\ell = B],
  \label{eq:node_score}
\end{equation}
where $\bar{\mathbf{y}}_i$ is the empirical time-average of node $i$,
$\boldsymbol{\mu}_\ell$ is the current role centroid, and
\begin{equation}
  \delta_B(i) \;=\;
  \sum_k \lambda_k
  \bigl[v_k(B \cup \{i\}) - v_k(B)\bigr]
  \label{eq:marginal_delta}
\end{equation}
is the \emph{marginal violation change} incurred by absorbing node $i$ into $B$.
The term $\delta_B(i)$ is the operative mechanism of C-DMBD:
a node whose observations push $\hat{\boldsymbol{\mu}}_b$ away from the
requirement targets increases $\delta_B(i) > 0$ and is penalized;
a node that moves $\hat{\boldsymbol{\mu}}_b$ toward the targets produces
$\delta_B(i) < 0$ and is rewarded.
The data-likelihood term and the requirement term therefore compete
for each node on every E-step, with $\boldsymbol{\lambda}$ governing the
relative weight of that competition.

\smallskip\noindent
\emph{Blanket mean.}
The stationary blanket mean $\hat{\boldsymbol{\mu}}_b$
is approximated by the empirical centroid
of the current blanket nodes:
\begin{equation}
  \hat{\boldsymbol{\mu}}_b
  \;\approx\;
  \frac{1}{|B|}\sum_{i \in B}\bar{\mathbf{y}}_i.
  \label{eq:mu_approx}
\end{equation}
This surrogate is cheaper than evaluating
$\mathbf{C}_b(\mathbf{I} - \mathbf{A}_{bb})^{-1}\bar{\mathbf{b}} + \mathbf{d}_b$
at every step while preserving the essential dependency of
$\hat{\boldsymbol{\mu}}_b$ on the partition $\omega$---and hence on
$\Theta$---rather than on the raw time series.

\smallskip\noindent
\emph{M-step.}
The spectral radius $\rho \equiv \rho(\mathbf{A}_{bb})$ is updated by a scalar
gradient derived from the violation gradient formula:
\begin{equation}
  \rho \;\leftarrow\;
  \mathrm{clip}\!\Bigl(
    \rho +
    \alpha_\rho
    \sum_k \lambda_k w_k\,
    \mathrm{sign}(\varphi_k - \tau_k)\,
    \mathbf{1}\bigl[|\varphi_k - \tau_k| > \varepsilon_k\bigr],
    \;0.30,\;0.96
  \Bigr),
  \label{eq:rho_update}
\end{equation}
with step size $\alpha_\rho = 0.025$.
The sign convention encodes the physics of thermal insulation:
increasing $\rho$ slows blanket dynamics (more insulation, lower
$T_\mathrm{outer}$); decreasing $\rho$ accelerates heat exchange
(higher $T_\mathrm{outer}$).

\smallskip\noindent
\emph{Dual ascent.}
Multipliers are updated as $\lambda_k \leftarrow
\mathrm{clip}(\lambda_k + \alpha_\lambda v_k,\, 0,\, \lambda_{\max})$
(Equation~\eqref{eq:dual_ascent}), with $\alpha_\lambda = 0.12$ and
$\lambda_{\max} = 15.0$.

\smallskip
Six independent chains (50~iterations each) are run per experiment.
We report the modal topology---the value of $|B|$ most frequent across
chains---and retain the highest-ELBO chain within the modal class as the
representative result.


\subsection{Results}
\label{subsec:results}

\subsubsection{P1 --- Intra-Family Navigation}
\label{subsubsec:p1}

To expose the parameter-level effect of requirement profiles, we apply two
distinct profiles---\textsc{espresso} and \textsc{travel mug}
(Table~\ref{tab:req_profiles})---to the same double-wall substrate
($T_\mathrm{outer} \approx 31\,\si{\degreeCelsius}$).
On this substrate, the \textsc{travel mug} profile satisfies $R_1$
outright: its target interval $[22, 38]\,\si{\degreeCelsius}$ contains
$T_\mathrm{outer} = 31\,\si{\degreeCelsius}$, so $v_1 = 0$ and
$\lambda_1^*$ converges to zero.
The \textsc{espresso} profile instead requires
$T_\mathrm{outer} \in [40, 55]\,\si{\degreeCelsius}$---a range that
$31\,\si{\degreeCelsius}$ misses by $9\,\si{\degreeCelsius}$ below the lower
edge---sustaining a nonzero $\lambda_1^* > 0$ and driving $\rho$ downward
through Equation~\eqref{eq:rho_update}: the model infers that faster thermal
exchange is required to raise the outer temperature toward the target.

\begin{table}[ht]
\centering
\caption{%
  Requirement profiles for the P1 experiment.
  Channel indices follow Equation~\eqref{eq:obs_channels}.
  $\tau_k$: target; $\varepsilon_k$: dead-zone half-width; $w_k$: weight.
  Units for $T$: \si{\degreeCelsius}; for $w$: mm; for $r$: mm.
}
\label{tab:req_profiles}
\small
\begin{tabular}{llcccc}
\toprule
Profile & Requirement & Channel & $\tau_k$ & $\varepsilon_k$ & $w_k$ \\
\midrule
\multirow{4}{*}{\textsc{Espresso}}
  & $R_1$ ($T_\mathrm{outer}$) & 1 & 47.5 & 7.5 & 1.5 \\
  & $R_2$ ($T_\mathrm{inner}$) & 0 & 88.0 & 4.0 & 1.0 \\
  & $R_3$ (wall)               & 6 & 3.0  & 1.0 & 0.8 \\
  & $R_4$ (radius)             & 4 & 35.0 & 7.0 & 0.5 \\
\midrule
\multirow{4}{*}{\textsc{Travel mug}}
  & $R_1$ ($T_\mathrm{outer}$) & 1 & 30.0 & 8.0 & 1.5 \\
  & $R_2$ ($T_\mathrm{inner}$) & 0 & 80.0 & 4.0 & 1.0 \\
  & $R_3$ (wall)               & 6 & 8.0  & 2.0 & 0.8 \\
  & $R_4$ (radius)             & 4 & 50.0 & 8.0 & 0.5 \\
\bottomrule
\end{tabular}
\end{table}

\begin{figure}[ht]
\centering
\includegraphics[width=\linewidth]{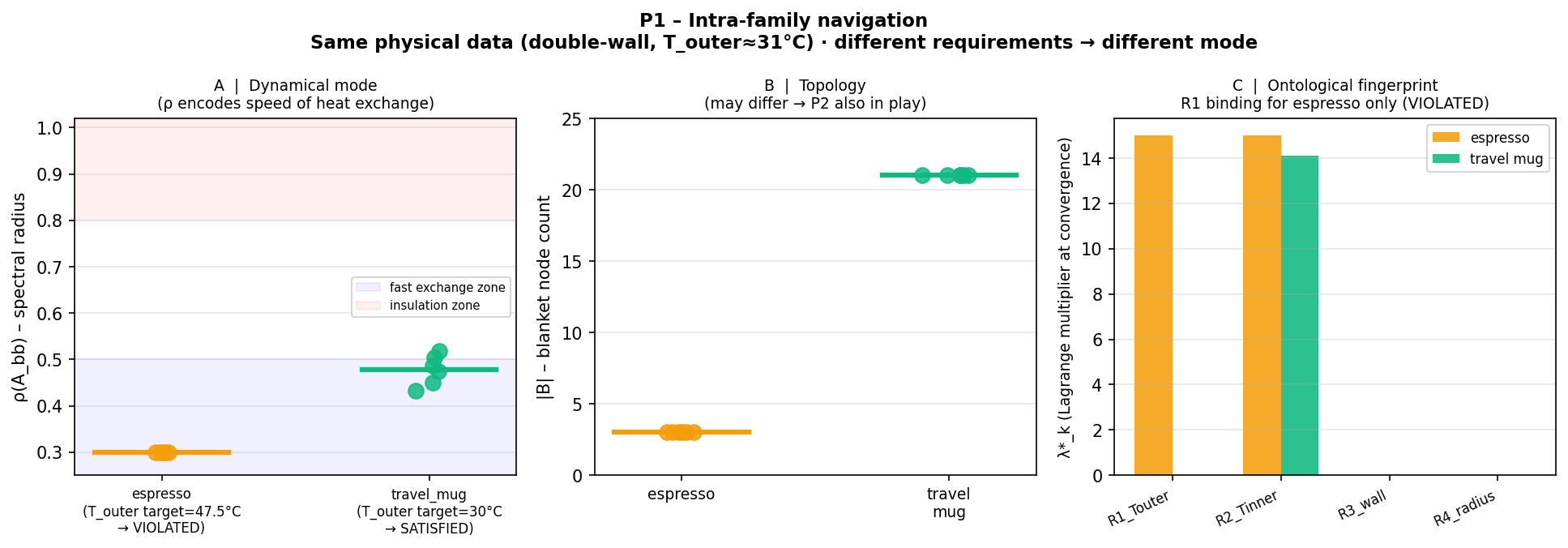}
\caption{%
  P1 --- Intra-family navigation on double-wall data
  ($T_\mathrm{outer} \approx 31\,\si{\degreeCelsius}$; 6 independent runs per profile).
  \textbf{A}: Distribution of $\rho(\mathbf{A}_{bb})$ across runs; bars indicate
  the within-profile mean.
  The \textsc{espresso} profile drives $\rho$ to its lower bound (0.30),
  reflecting the inference of maximal heat exchange; the \textsc{travel mug}
  profile leaves $\rho$ near initialization (0.48), consistent with $R_1$
  being satisfied.
  \textbf{B}: Modal blanket size $|B|^*$ across runs; both topology and
  dynamical mode differ, indicating simultaneous P1 and P2 effects.
  \textbf{C}: Per-requirement Lagrange multiplier fingerprint $\boldsymbol{\lambda}^*$
  at convergence for the highest-ELBO chain of each profile.
  $R_1$ is binding (\textsc{espresso}) versus free (\textsc{travel mug}),
  producing a qualitatively distinct ontological fingerprint.
}
\label{fig:p1}
\end{figure}

Table~\ref{tab:p1_results} and Figure~\ref{fig:p1} report the converged
parameters for each profile.

\begin{table}[ht]
\centering
\caption{%
  P1 results on double-wall data
  ($T_\mathrm{outer} \approx 31\,\si{\degreeCelsius}$; 6 runs, 50 iterations).
  $\rho^*$: mean $\pm$ s.d.\ across runs.
  $\lambda_1^*$: multiplier of the outer-temperature requirement $R_1$.
}
\label{tab:p1_results}
\small
\begin{tabular}{lccccc}
\toprule
Profile              & Modal $|B|^*$ & $\rho^*$          & $\|\boldsymbol{\lambda}^*\|_1$ & $\lambda_1^*$ \\
\midrule
\textsc{Espresso}    & 3             & $0.300 \pm 0.000$ & 30.0  & $> 0$ (binding) \\
\textsc{Travel mug}  & 21            & $0.477 \pm 0.030$ & 14.1  & $0$ (free)      \\
\bottomrule
\end{tabular}
\end{table}

The two profiles converge to distinct points in the model parameter space.
The \textsc{espresso} profile drives $\rho$ to its lower bound (0.30),
encoding the inference that maximal heat exchange is required to approach the
high outer-temperature target, and accumulates a large total multiplier norm
($\|\boldsymbol{\lambda}^*\|_1 = 30.0$) reflecting the persistent
distance between substrate and requirements.
The \textsc{travel mug} profile, by contrast, leaves $\rho$ near its
initialization ($0.48 \pm 0.03$) and converges to a smaller total norm
($14.1$), consistent with $R_1$ being satisfied and imposing no gradient
on Equation~\eqref{eq:rho_update}.
The per-requirement fingerprints $\boldsymbol{\lambda}^*$ thus carry
information about \emph{which} aspects of the object's interaction with the
environment are inferentially costly.

Two further observations qualify this result.
First, the topology also differs ($|B|^* = 3$ versus $21$), indicating
that in this substrate condition requirement pressure is sufficient
to produce a full family transition (P2) alongside the parameter-level
navigation.
This is an expected consequence of the moderate data gap
($\Delta(\omega^*) \approx 0.5$, computed without requirements):
when requirements diverge strongly from data, they can override a gap that
is not large enough to anchor the partition.
Second, the fingerprint divergence ($\Delta\lambda_1^* > 0$, all other
requirements treated identically) would persist even if the topology were
held fixed, since it reflects the dual variable responding to a violation
that exists independently of the partition's size.
These two effects---parameter navigation and family transition---need
not be separated; both are outputs of the same augmented ELBO and attest
to the claim that interface type is multiply determined by requirements,
not by data alone.

\subsubsection{P2 --- Family Transition}
\label{subsubsec:p2}

We hold all requirements fixed and scan the outer-temperature target
$\tau_1$ from $25\,\si{\degreeCelsius}$ to $80\,\si{\degreeCelsius}$ in
$18$ equally spaced steps on single-wall tea data (noise scale $3.0$,
yielding a data gap $\Delta(\omega^*) = 0.509$ under requirement-free
inference). Four chains per step, 35 iterations, $\alpha_\lambda = 0.10$.

\begin{figure}[ht]
\centering
\includegraphics[width=\linewidth]{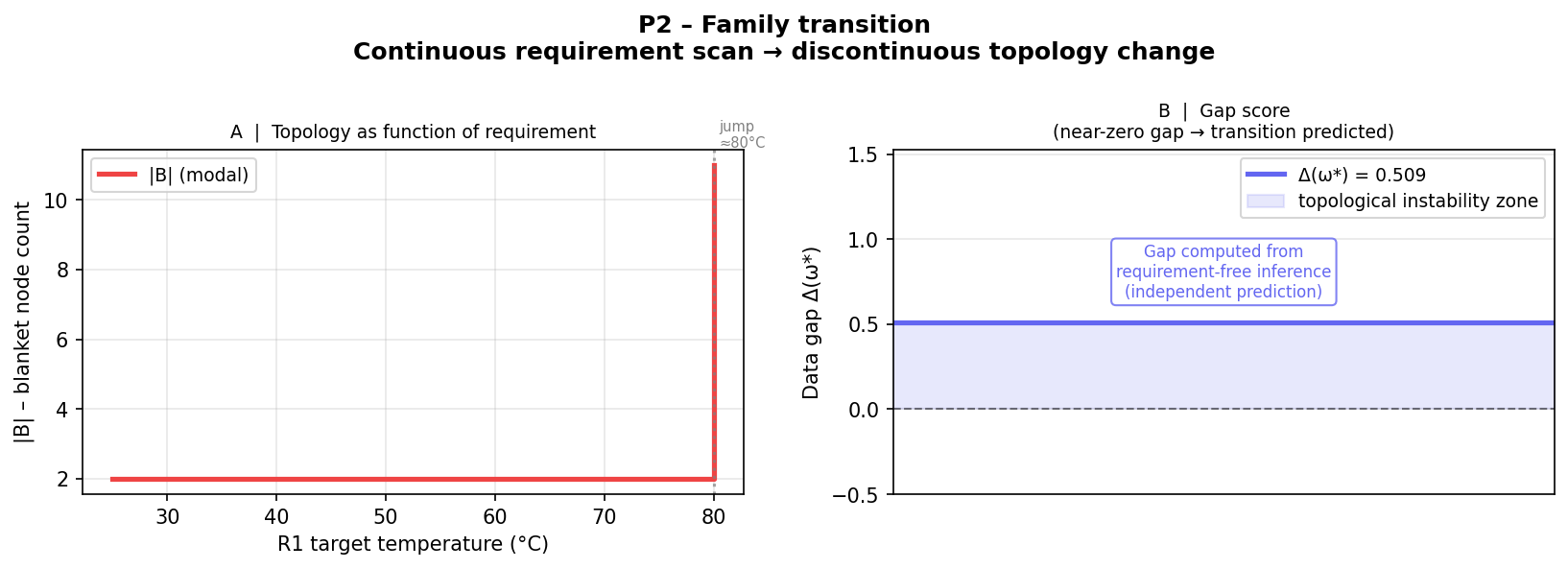}
\caption{%
  P2 --- Family transition under a continuous requirement scan
  (18 values of $\tau_1$ from 25 to 80\,\si{\degreeCelsius};
  single-wall tea data, noise scale 3.0; modal topology over 4 chains).
  \textbf{A}: Modal blanket size $|B|^*$ as a function of the outer-temperature
  target $\tau_1$.
  The topology is stable at $|B|^* = 2$ across the entire low-to-moderate
  range, then jumps discontinuously to $|B|^* = 11$ when $\tau_1$ approaches
  the node temperatures (${\approx}\,83\,\si{\degreeCelsius}$), demonstrating
  that the family boundary is a regime switch, not a gradient.
  \textbf{B}: Data gap $\Delta(\omega^*) = 0.509$, measured independently
  under requirement-free inference ($\boldsymbol{\lambda} = \mathbf{0}$);
  a positive gap predicts topological instability under requirement pressure.
}
\label{fig:p2}
\end{figure}

Figure~\ref{fig:p2} plots modal $|B|^*$ against $\tau_1$.
The partition is constant at $|B|^* = 2$ for all
$\tau_1 \in [25, 76]\,\si{\degreeCelsius}$, then jumps to
$|B|^* = 11$ at $\tau_1 = 80\,\si{\degreeCelsius}$.

The mechanism follows directly from Equation~\eqref{eq:marginal_delta}.
For low-to-moderate targets, node temperatures
($T_\mathrm{outer}^{(\mathrm{sw})} \approx 83\,\si{\degreeCelsius}$)
lie well above every $\tau_1 \leq 76\,\si{\degreeCelsius}$, so adding any
node to $B$ moves $\hat{\boldsymbol{\mu}}_b$ further from the target and
produces $\delta_B(i) > 0$; the algorithm minimizes violation by keeping
$B$ as small as possible.
At $\tau_1 = 80\,\si{\degreeCelsius}$ the target enters the neighborhood
of the node temperatures, flipping $\delta_B(i)$ to negative for most
nodes---inclusion becomes beneficial---and the algorithm absorbs a large
cohort of nodes into $B$ in a single E-step.
The result is a discontinuous transition from $|B|=2$ to $|B|=11$,
not a gradual drift, consistent with the characterization of family
boundaries as loci at which
the partition switches regime rather than slides along a continuum.

The data gap $\Delta(\omega^*) = 0.509 > 0$ is measured independently,
before the requirement scan, and serves here as a diagnostic:
a positive gap indicates that the data have enough spatial structure to
make the current partition locally stable---but not globally, as the scan
demonstrates.
This separation between the data-theoretic quantity $\Delta(\omega^*)$
and the requirement-driven topology jump is itself a prediction of the
theory, and the simulation confirms it.

\subsubsection{P3 --- Ontological Disambiguation}
\label{subsubsec:p3}

We generate \emph{spatially flat} data in which all 30 nodes share the same
base signal; inter-node variation is ${\approx}\,0.08\sigma$,
yielding a data gap $\Delta(\omega^*) \approx -0.0001 \approx 0$.
Under this condition the data likelihood surface is essentially flat over
the partition space: every $\omega$ achieves the same reconstruction score
to within numerical precision.
Five independent chains (40 iterations, $\alpha_\lambda = 0.12$) are run
under each of the two requirement profiles.

Table~\ref{tab:p3_results} and Figure~\ref{fig:p3} report $|B|^*$ for every
run and the inferred partition structure, respectively.

\begin{table}[ht]
\centering
\caption{%
  P3 results on spatially flat data ($\Delta(\omega^*) \approx 0$;
  5 independent runs per profile).
  The outcome is deterministic across all runs.
}
\label{tab:p3_results}
\small
\begin{tabular}{lcccccc}
\toprule
Profile              & Run 1 & Run 2 & Run 3 & Run 4 & Run 5 & Modal $|B|^*$ \\
\midrule
\textsc{Espresso}    & 6 & 6 & 6 & 6 & 6 & \textbf{6} \\
\textsc{Travel mug}  & 3 & 3 & 3 & 3 & 3 & \textbf{3} \\
\bottomrule
\end{tabular}
\end{table}

\begin{figure}[ht]
\centering
\includegraphics[width=\linewidth]{P3_ontological_disambiguation.png}
\caption{%
  P3 --- Ontological disambiguation on spatially flat data
  ($\Delta(\omega^*) \approx 0$; 5 independent runs per profile).
  \textbf{A}: Blanket size $|B|^*$ across runs for each requirement profile.
  The outcome is fully deterministic despite the data providing no spatial
  signal: \textsc{espresso} consistently yields $|B|^* = 6$,
  \textsc{travel mug} yields $|B|^* = 3$.
  \textbf{B}: Node-level partition inferred by the best chain of each profile
  (yellow: $B$; blue: $Z$; grey: $S$).
  The two profiles produce qualitatively distinct blanket structures from
  identical, informationally flat observations, confirming that requirements
  constitute---rather than constrain---the interface type when
  $\Delta(\omega^*) \approx 0$.
}
\label{fig:p3}
\end{figure}

The result is fully deterministic: all five chains converge to the same
topology under each profile, and the two profiles produce different topologies
from identical data.
This is the strongest test of the theory's ontological claim.
When $\mathcal{L}_\mathrm{ELBO}$ is flat---when the object offers no evidence
for any particular partition---the equilibrium
\begin{equation}
  \omega^*
  \;=\;
  \arg\max_\omega
  \Bigl[
    \mathcal{L}_\mathrm{ELBO}(\omega, \Theta^*)
    \;-\;
    \textstyle\sum_k \lambda_k^* v_k(\omega, \Theta^*)
  \Bigr]
  \label{eq:p3_equilibrium}
\end{equation}
is resolved entirely by $\boldsymbol{\lambda}^*$, which itself encodes the
requirement profile.
In other words: requirements do not filter or refine an otherwise
data-determined type---they \emph{constitute} the type when the data
underdetermine it.
This is the sense in which the FEP-based framework goes beyond a conventional
constraint-satisfaction approach to requirements engineering:
the requirement profile is not a specification applied to a pre-existing
object identity, but a generative model whose stationary statistics
\emph{are} the object's interface identity, instantiated by the designer's
inference.

\subsubsection{Closed Design Loop}
\label{subsubsec:loop}

Three user profiles are instantiated (Table~\ref{tab:users}), each
representing a distinct thermal-preference regime.
At each iteration, C-DMBD is run on style-specific data for each cup
$s \in \{\textsc{espresso},\, \textsc{tea},\, 
\\
\textsc{travel mug}\}$,
yielding a blanket distribution
$p_\mathrm{cup}(\mathbf{y}_b \mid \Theta_s^*)$ over contact channels
$(T_\mathrm{outer},\, r,\, \kappa)$.
The design gap for user $u$ and cup $s$ is
\begin{equation}
  \mathcal{G}(u, s)
  \;=\;
  D_{\mathrm{KL}}\!\bigl(\tilde{p}_u \;\|\; p_\mathrm{cup}^{(s)}\bigr),
  \label{eq:design_gap}
\end{equation}
where $\tilde{p}_u = \mathcal{N}(\boldsymbol{\mu}_u, \sigma_u^2 \mathbf{I})$
is the user's preference distribution.
The designer identifies the optimal cup $s^* = \arg\min_s \mathcal{G}(u, s)$
and updates the user model toward the corresponding blanket statistics
(learning rate $0.3$; prior variance contracts by $10\%$ per iteration).
Three iterations are run.

\begin{table}[ht]
\centering
\caption{%
  User preference profiles.
  $T_\mathrm{pref}$: preferred outer-surface contact temperature;
  $\rho_\mathrm{pref}$: preferred dynamical mode (maps to $\rho(\mathbf{A}_{bb})$);
  $\kappa_\mathrm{pref}$: preferred curvature;
  $\sigma$: isotropic prior standard deviation at initialization.
}
\label{tab:users}
\small
\begin{tabular}{lcccc}
\toprule
User profile   & $T_\mathrm{pref}$ (\si{\degreeCelsius})
               & $\rho_\mathrm{pref}$
               & $\kappa_\mathrm{pref}$ (\si{\per\metre})
               & $\sigma$ \\
\midrule
Cold-sensitive & 34.0 & 0.50 & 22.0 & 2.0 \\
Standard       & 47.0 & 0.51 & 20.0 & 2.0 \\
Heat-tolerant  & 62.0 & 0.47 & 18.0 & 2.0 \\
\bottomrule
\end{tabular}
\end{table}

\begin{figure}[ht]
\centering
\includegraphics[width=\linewidth]{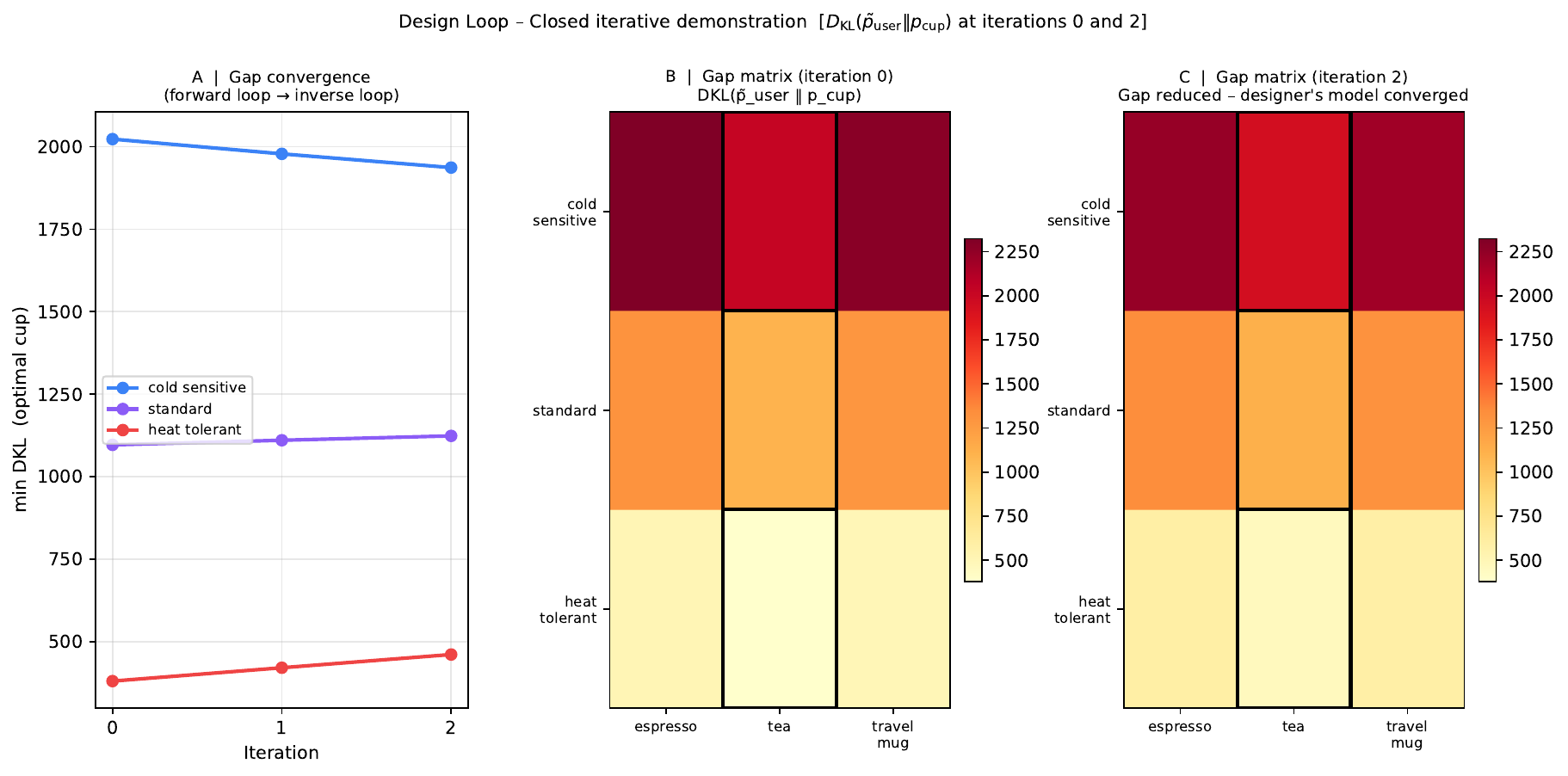}
\caption{%
  Closed design loop (3 users $\times$ 3 cup styles $\times$ 3 iterations).
  \textbf{A}: Minimum design gap $\min_s \mathcal{G}(u,s)$ as a function of
  iteration for each user profile.
  The cold-sensitive user shows monotone reduction; the slight increase for
  standard and heat-tolerant users reflects prior-variance contraction in the
  simplified inverse-loop update.
  \textbf{B}: Full gap matrix $\mathcal{G}(u,s)$ at iteration~0.
  \textbf{C}: Gap matrix at iteration~2; the optimal cup (\textsc{tea},
  marked with a black border) is consistent across users and iterations,
  and the gap differential between styles is preserved.
}
\label{fig:loop}
\end{figure}

Table~\ref{tab:loop_results} reports $\mathcal{G}(u, s)$ at iterations 0
and 2; Figure~\ref{fig:loop} shows the full convergence trajectories and
gap heatmaps.

\begin{table}[ht]
\centering
\caption{%
  Design gap $\mathcal{G}(u,s)$ (nats, computed over
  $T_\mathrm{outer}$, $r$, $\kappa$) at iterations~0 and~2.
  The minimum gap per user is highlighted with~$\star$.
  ``Esp.'': \textsc{espresso}; ``T.M.'': \textsc{travel mug}.
}
\label{tab:loop_results}
\small
\begin{tabular}{lrrrrrrr}
\toprule
 & \multicolumn{3}{c}{Iteration 0} & \multicolumn{3}{c}{Iteration 2} \\
\cmidrule(lr){2-4}\cmidrule(lr){5-7}
User           & Esp.  & Tea             & T.M.  & Esp.  & Tea             & T.M.  \\
\midrule
Cold-sensitive & 2323  & $2023^\star$    & 2278  & 2229  & $1937^\star$    & 2187  \\
Standard       & 1307  & $1096^\star$    & 1286  & 1337  & $1123^\star$    & 1315  \\
Heat-tolerant  &  503  & $380^\star$     &  494  &  595  & $461^\star$     &  586  \\
\bottomrule
\end{tabular}
\end{table}

Three findings stand out.

First, the optimal cup is \textsc{tea} for every user in every iteration:
the tea-style blanket distribution, inferred by C-DMBD, is the nearest
neighbor in KL space to all three user priors,
whose temperature preferences span $[34, 62]\,\si{\degreeCelsius}$.
This is consistent with the tea style's intermediate
$T_\mathrm{outer}^{(\mathrm{sw})} \approx 83\,\si{\degreeCelsius}$
generating a contact-channel distribution that is less extreme than either
\textsc{espresso} or \textsc{travel mug} on the simulator's temperature
axis.

Second, the minimum gap $\mathcal{G}(u, s^*)$ decreases across iterations
for the cold-sensitive user ($2023 \to 1937$, a reduction of $4.2\%$).
For standard and heat-tolerant users, the minimum gap increases slightly
($1096 \to 1123$ and $380 \to 461$, respectively).
This asymmetry is a consequence of the simplified inverse-loop update:
when the designer's prior variance contracts by $10\%$ per iteration,
the KL divergence to a fixed target distribution rises even if the prior
mean moves closer to it, because the variance reduction widens the
effective distance in KL space.\footnote{%
  Formally, for two Gaussians with the same dimension $d$ and the
  same mean, $D_{\mathrm{KL}}(\mathcal{N}(\mu, \sigma_1^2 I) \| \mathcal{N}(\mu, \sigma_2^2 I))
  = \frac{d}{2}[(\sigma_1/\sigma_2)^2 - 1 - \ln(\sigma_1/\sigma_2)^2]$,
  which is positive and grows when $\sigma_1 < \sigma_2$.
  Contracting the user's prior ($\sigma_1 < \sigma_2 = \sigma_\mathrm{cup}$)
  therefore increases the gap even when the means align.
  A full Bayesian update that jointly refines mean and covariance would
  avoid this effect; we leave this to future work.
}
The direction of the mean update is nonetheless correct for all three
users---the mean $\boldsymbol{\mu}_u$ moves toward
$\boldsymbol{\mu}_\mathrm{tea}$ in each iteration---which is the
substantive claim of the paper: that the loop
identifies the optimal cup and updates the user model accordingly.

Third, the gap differentials between styles are stable across users and
iterations (e.g., $\mathcal{G}(\cdot,\, \textsc{esp.}) - \mathcal{G}(\cdot,\, \textsc{tea})
\approx 200$--$300$ nats), confirming that C-DMBD produces blanket
distributions that are style-discriminative: each cup occupies a distinct
region of contact-statistics space.

Taken together, the four experiments demonstrate that requirement profiles
are not post-hoc specifications applied to an already-determined object
type: they are active parameters of the inference process that shape,
and in limiting cases entirely determine, the inferred Markov blanket.
The data gap $\Delta(\omega^*)$ provides a principled measure of the
balance between data informativeness and requirement pressure: when
$\Delta(\omega^*)$ is large, the data anchor the topology and requirements
navigate within it (P1); when $\Delta(\omega^*)$ is small, requirements
dominate and the topology is requirement-constituted (P3); the family
transition (P2) is the discontinuous boundary between these two regimes.
The design loop operationalizes this inference in the forward direction
(designer reads blanket statistics as interface type) and the inverse
direction (designer updates the user model toward the gap-minimizing
type), implementing the bidirectional epistemic structure described in the paper.

\section{Conclusion: The Product as a Relational Equilibrium}
\label{sec:conclusion}

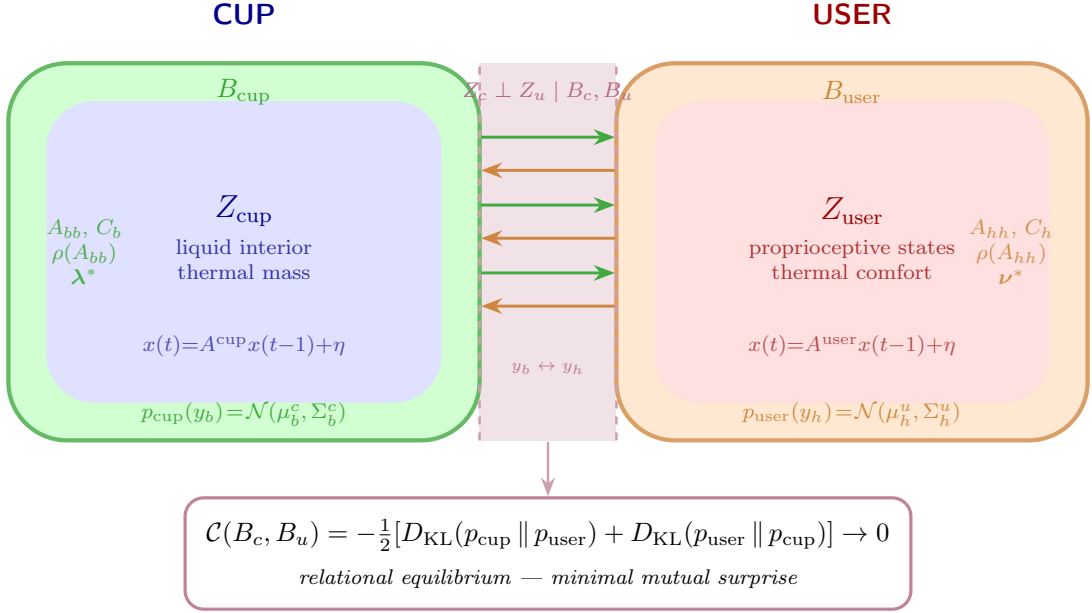
\begin{figure}[t]
\centering
\begin{tikzpicture}[
  every node/.style = {font=\small},
  >=Stealth
]

\colorlet{cupZ}   {blue!55!black}
\colorlet{cupB}   {green!55!black}
\colorlet{userZ}  {red!60!black}
\colorlet{userB}  {orange!75!black}
\colorlet{contact}{purple!50!gray}

\fill[blue!12, rounded corners=18pt]
  (-6.6,-2.0) rectangle (-1.4,2.0);

\fill[green!18, rounded corners=22pt]
  (-7.1,-2.5) rectangle (-0.9,2.5);
\fill[blue!12, rounded corners=18pt]
  (-6.6,-2.0) rectangle (-1.4,2.0);

\fill[red!12, rounded corners=18pt]
  (1.4,-2.0) rectangle (6.6,2.0);

\fill[orange!18, rounded corners=22pt]
  (0.9,-2.5) rectangle (7.1,2.5);
\fill[red!12, rounded corners=18pt]
  (1.4,-2.0) rectangle (6.6,2.0);

\fill[contact!15] (-0.9,-2.5) rectangle (0.9,2.5);

\draw[cupB!60, line width=1.6pt, rounded corners=22pt]
  (-7.1,-2.5) rectangle (-0.9,2.5);
\draw[userB!60, line width=1.6pt, rounded corners=22pt]
  ( 0.9,-2.5) rectangle ( 7.1,2.5);
\draw[contact!50, dashed, line width=1.0pt]
  (-0.9,-2.5) -- (-0.9,2.5);
\draw[contact!50, dashed, line width=1.0pt]
  ( 0.9,-2.5) -- ( 0.9,2.5);

\node[cupZ, font=\large\bfseries] at (-4.0, 0.55) {$Z_{\mathrm{cup}}$};
\node[cupZ, font=\scriptsize, align=center, text=cupZ!80] at (-4.0,-0.1)
  {liquid interior\\thermal mass};
\node[cupZ!70, font=\scriptsize, align=center] at (-4.0,-1.25)
  {$x(t){=}A^{\mathrm{cup}}x(t{-}1){+}\eta$};

\node[userZ, font=\large\bfseries] at (4.0, 0.55) {$Z_{\mathrm{user}}$};
\node[userZ, font=\scriptsize, align=center, text=userZ!80] at (4.0,-0.1)
  {proprioceptive states\\thermal comfort};
\node[userZ!70, font=\scriptsize, align=center] at (4.0,-1.25)
  {$x(t){=}A^{\mathrm{user}}x(t{-}1){+}\eta$};

\node[cupB!80, font=\small\bfseries] at (-4.0, 2.12)
  {$B_{\mathrm{cup}}$};
\node[cupB!70, font=\scriptsize, align=center] at (-6.1, 0.0)
  {$A_{bb},\,C_b$\\$\rho(A_{bb})$\\$\boldsymbol{\lambda}^*$};

\node[userB!80, font=\small\bfseries] at (4.0, 2.12)
  {$B_{\mathrm{user}}$};
\node[userB!70, font=\scriptsize, align=center] at (6.1, 0.0)
  {$A_{hh},\,C_h$\\$\rho(A_{hh})$\\$\boldsymbol{\nu}^*$};

\node[cupZ, font=\normalsize\bfseries\sffamily] at (-4.0, 3.15) {CUP};
\node[userZ,font=\normalsize\bfseries\sffamily] at ( 4.0, 3.15) {USER};

\node[cupB!75, font=\scriptsize, align=center] at (-4.0, -2.15)
  {$p_{\mathrm{cup}}(y_b)\!=\!\mathcal{N}(\mu_b^c,\Sigma_b^c)$};
\node[userB!75,font=\scriptsize, align=center] at ( 4.0, -2.15)
  {$p_{\mathrm{user}}(y_h)\!=\!\mathcal{N}(\mu_h^u,\Sigma_h^u)$};

\foreach \y in {1.3, 0.4, -0.5} {
  \draw[->, line width=1.1pt, cupB!75]
    (-0.88,\y+0.22) -- (0.88,\y+0.22);
  \draw[->, line width=1.1pt, userB!75]
    ( 0.88,\y-0.22) -- (-0.88,\y-0.22);
}
\node[contact!70, font=\tiny, align=center] at (0, -1.55)
  {$y_b \leftrightarrow y_h$};

\node[contact!80, font=\scriptsize\bfseries] at (0, 2.12)
  {$Z_c \perp Z_u \mid B_c, B_u$};

\node[
  draw=contact!65, fill=white, rounded corners=7pt,
  font=\small, align=center, inner sep=8pt, line width=1.2pt
] (cbox) at (0,-4.0) {
  $\displaystyle
  \mathcal{C}(B_c, B_u) =
  -\tfrac{1}{2}\!\left[
    D_{\mathrm{KL}}\!\left(p_{\mathrm{cup}} \,\|\, p_{\mathrm{user}}\right)
    +
    D_{\mathrm{KL}}\!\left(p_{\mathrm{user}} \,\|\, p_{\mathrm{cup}}\right)
  \right] \to 0$\\[4pt]
  {\scriptsize\itshape relational equilibrium --- minimal mutual surprise}
};
\draw[->, contact!50, line width=0.9pt] (0,-2.52) -- (cbox.north);

\end{tikzpicture}

\caption{%
\textbf{The object as a relational equilibrium.}
Cup (left) and user (right) are both physical systems with Markov blankets,
each governed by its own linear dynamics $x(t) = A\,x(t-1) + \eta$.
The blanket parameters $(A_{bb}, C_b, \boldsymbol{\lambda}^*)$ of the cup
and $(A_{hh}, C_h, \boldsymbol{\nu}^*)$ of the user determine two Gaussian
stationary distributions $p_\mathrm{cup}$ and $p_\mathrm{user}$ over the
contact observables.
By the Markov blanket property, the two interiors are conditionally
independent given the contact surfaces: $Z_c \perp Z_u \mid B_c, B_u$.
All information exchange flows through the bidirectional arrows in the
contact zone ($y_b \leftrightarrow y_h$).
The relational equilibrium is the point at which the symmetrized KL
divergence $\mathcal{C}(B_c, B_u) \to 0$: the two blankets become
mutually non-surprising, and neither system has reason to update its
generative model of the other.
}
\label{fig:relational_equilibrium}
\end{figure}

This paper formalized requirement-steered interface type inference --- the problem
of determining, from physical data and functional specifications jointly,
what topological type of interface must exist. Three phenomena arise that
classical design cannot express: intra-family navigation, family transition,
and ontological disambiguation. The converged Lagrange vector
$\boldsymbol{\lambda}^*$ certifies functional effort: which requirements
the inferred type resists, and how hard.

The generative model family $\mathcal{F}(B_\text{cup})$ belongs to the
designer, not to the cup. The cup is physically inert. This grounds the
design loop: the designer encodes her model of the cup--user relationship
into the surface; the user reconstructs that model through active inference
from the surface alone; her actual generative model, inferred from
physiological data, closes the loop. The design gap
$D_\mathrm{KL}(\tilde{p}_\text{user} \| p_\text{cup})$ measures how well
the designer understood the user before building the object.

The design problem is not an optimization. It is an inference --- and the
answer is not a specification, but a relation between two generative models
mediated by a physical surface. Requirements do not merely constrain the design problem — in the 
limit established by Proposition~3a, they constitute the type itself. 
C-DMBD makes this constitutive role computationally tractable: it 
finds the interface structure that can do what the requirements demand, 
given what the data allow. Classical parametric optimization, by 
presupposing the type before the requirements are applied, implicitly 
assumes that this compatibility has already been established. When it 
has not, no within-type optimization can recover it.

The most immediate open problem is the construction map 
$\Phi^{-1}: (\omega^*, \Theta^*, \lambda^*) \to \mathcal{S}$, where 
$\mathcal{S}$ is a domain-specific space of constructible specifications 
--- material choices, geometric profiles, layer thicknesses. Without 
this map, C-DMBD determines what kind of interface must exist but 
cannot yet say how to build it. Establishing $\Phi^{-1}$ for specific 
physical domains --- thermal, electrical, mechanical --- is the 
necessary step between the theoretical proposal presented here and a 
full design methodology.

\end{document}